\documentclass[11pt]{JHEP3}
\usepackage{graphicx,amsmath,amssymb}

\def\bk{{\bf k}}

\def\CO{{\cal O}}

\def\ap{\alpha'}

\def\ap{{\alpha'}}

\def\half{\frac{1}{2}}

\newcommand{\Mpl}{M_\mathrm{pl}}

\newcommand{\be}{\begin{equation}}
\newcommand{\ee}{\end{equation}}
\newcommand{\bea}{\begin{eqnarray}}
\newcommand{\eea}{\end{eqnarray}}
\newcommand{\barr}{\begin{array}}
\newcommand{\earr}{\end{array}}
\newcommand{\tileta}{{\tilde \eta}}
\newcommand{\calO}{{\cal O}}
\newcommand{\calR}{{\cal R}}

\title{Duality Cascade in Brane Inflation}
\author{ Rachel Bean$^{1}$\footnote{rbean@astro.cornell.edu}~, 
Xingang Chen$^{2,3}$\footnote{xgchen@mit.edu}~,
Girma Hailu$^{4}$\footnote{hailu@lepp.cornell.edu}~,
S.-H. Henry Tye$^{3,4}$\footnote{tye@lepp.cornell.edu}~ 
and Jiajun Xu$^{3,4}$\footnote{jx33@cornell.edu} 
\\\small{\em $^1$Department of Astronomy, Cornell University, Ithaca, NY
14853, USA \\
$^2$Center for Theoretical Physics, \\
Massachusetts Institute of Technology, Cambridge, MA 02139, USA \\
$^3$Kavli Institute for Theoretical Physics in China, \\
Chinese Academy of Science, Beijing 100080, P.R.China \\
$^4$Newman Laboratory for Elementary Particle Physics, \\
Cornell University, Ithaca, NY 14853, USA} }

\abstract{
We show that brane inflation is very sensitive to tiny sharp features
in extra dimensions, including those in the potential and in the warp factor.
This can show up as observational signatures in the power spectrum
and/or non-Gaussianities of the cosmic microwave 
background radiation (CMBR).
One general example of such sharp features is a succession of small steps
in a warped throat, caused by Seiberg duality cascade using gauge/gravity duality.
We study the cosmological observational consequences of these steps in brane inflation.
Since the steps come in a series, the prediction of other steps
and their properties can be tested by future data and analysis.
It is also possible that the steps are too close to be resolved in the power spectrum,
in which case they may show up only in the non-Gaussianity of the CMB temperature fluctuations and/or EE polarization.
We study two cases.
In the slow-roll scenario where steps appear in the inflaton
potential, the sensitivity of brane inflation to the height and width of the steps
is increased by several orders of magnitude comparing to that in
previously studied large field models.
In the IR DBI scenario where steps appear in the warp factor, we find
that the glitches in the power spectrum caused by these sharp features are generally small
or even unobservable, but associated distinctive non-Gaussianity can be large.
Together with its large negative running of the power spectrum index, this scenario
clearly illustrates how rich and different a brane inflationary
scenario can be when
compared to generic slow-roll inflation. Such distinctive stringy
features may provide a powerful probe of superstring theory.
}

\preprint{MIT-CTP-3921 \\ CAS-KITPC/ITP-046}

\begin{document}

\newpage

\section{Introduction}
 
It is generally believed that the early universe went through an inflationary epoch \cite{Guth:1980zm}.
In the field theory context, the most studied realization is the slow-roll inflationary scenario \cite{Linde}.
Such models confront primordial cosmological observation via two data points : the magnitude of
the density perturbation (or temperature fluctuation in the cosmic microwave background radiation (CMBR)) as measured by COBE \cite{Smoot:1992td} $\delta_H \sim 10^{-5}$, and the
approximately scale invariant scalar power spectrum, with index $n_{s}  \sim 1$.
In addition, there are three observational upper bounds on, namely,
the tensor to scalar ratio $r< 0.3$, the running of the spectral index $|d n_{s}/d \ln
k| < 0.1$ and the non-Gaussianity $f_{NL} < \calO(100)$ \cite{Hinshaw:2006ia,Spergel:2006hy}.
Typical predictions of $d n_{s}/d \ln k$ and $f_{NL}$ (and maybe even $r$) may very well be
far too small to be detected.
Since the COBE normalization typically fixes the scale of the model, so far, the deviation of $n_{s}$ from unity is the only parameter that discriminates among slow-roll inflationary models. It is not surprising, therefore, that present data allows a large degeneracy of
models \cite{Bean:2008ga}.

If superstring theory is correct, it must provide an inflationary
scenario for the early universe. It should provide a precise and
restrictive framework for inflationary model building while at the
same time, new scenarios and new phenomena are likely to appear. Hopefully it will
predict distinctive stringy signatures that, if observed, can be used
to support superstring theory. It is along this direction that this
paper attempts to address. Here we like to discuss the possibility of
a stringy feature that may show up in a distinctive fashion in the
CMBR. If the proposal is realized in nature, the number of data points
linking theory and cosmological observation may increase (from $\sim$
2) by as much as an order of magnitude, substantially strengthening
the test of inflation as well as the specific stringy realization of
inflation.

It has become clear that the brane inflationary scenario is quite
natural in string theory \cite{Tye:2006uv}. Here, the inflaton is
simply the position of a mobile brane in the compactified bulk
\cite{Dvali:1998pa,Burgess:2001fx}. 
In flux compactification in Type IIB string
theory, where all closed string 
moduli are dynamically stabilized, warped geometry
appears naturally in the compactified bulk
\cite{Giddings:2001yu}. Warped geometry in the 6-dimensional
compactified bulk tends to appear as warped throats. In the simple but
realistic scenarios, inflation takes place as $D$3-branes move up or
down a warped throat
\cite{Kachru:2003sx,Silverstein:2003hf,Chen:2004gc}. 
Inflation typically ends
when the $D$3-branes reach the bottom of a throat and annihilate
the anti$-D$3-branes sitting there. The energy released heats up the
universe and starts the hot big bang.

The warped geometry either shifts masses to the infrared and helps to
flatten the inflaton potential, or provides speed-limit that restricts
the inflaton velocity, thus improving the e-folds of
inflation. A typical throat is a warped deformed (or resolved)
conifold. A string scale test mass in the bulk is red-shifted to a
smaller value at the bottom of the throat. Via gauge/gravity duality,
moving down the throat corresponds to a gauge theory flowing towards
the infrared region. The best known example suitable for brane
inflation is the $\mathcal{N}=1$ supersymmetric Klebanov-Strassler
(KS) throat in type IIB string theory on approximate $AdS_5 \times
T^{1,1}$ background \cite{Klebanov:2000hb}. It is argued that the KS
throat is dual to the $\mathcal{N}=1$ supersymmetric $SU(N+M)\times
SU(N)$ (with $N=KM$) gauge theory with bifundamental chiral
superfields and a quartic tree level superpotential in four
dimensions \cite{Klebanov:2000nc,Dymarsky:2005xt}. As the gauge theory flows towards the infrared,
$SU(N+M)=SU((K+1)M)$ becomes strongly coupled and Seiberg
``electric-magnetic'' duality says that it is equivalent to a weakly
coupled $SU((K-1)M)$ with appropriate matter content
\cite{Seiberg:1994pq}. As a result, the gauge theory becomes
$SU((K-1)M)\times SU(KM)$ with the same set of bifundamentals. This
process repeats as the gauge theory continues its flow towards the
infrared; that is, it goes through a series of Seiberg duality
transitions. This is known as the duality cascade
\cite{Strassler:2005qs}.

 Recently, a more detailed analysis of this Seiberg duality cascade \cite{Hailu:2006uj}
shows that there are corrections to the anomalous mass dimensions and
their effects in the renormalization group flow of the
couplings on the gauge theory side show up as small structures on the supergravity side: as steps appearing in the warped metric. The steps in the warp factor are expected to be slightly smoothed out, leading to a cascade feature that may be observable in cosmology. 
It was conjectured in Ref.\cite{Hailu:2006uj} that the origin of the steps might be
localized charges (branes) at radial positions inside the throat. To fully
understand the gravity dual of the duality cascade, one needs to take the
branes into account and study the resulting supergravity solutions
together with the thickness and the stability of the branes including
other smooth corrections to the KS solution. We leave this
issue for future study and in this paper we focus only on the observable
effects due to the steps.
The presence of steps is quite generic: either as a correction to the approximate
geometry for a warped throat, or as a variation of the simplest warped deformed throat
in the KKLMMT-like scenario.
Although the KS throat is the best understood example, in some sense it is too simple, i.e., there are no chiral fields in the infrared. Warped throats are natural consequences of flux compactification. They arise
by considering Calabi-Yau singularities that admit complex deformations, corresponding to smoothing the singular points with 3 cycles on which to turn on fluxes. The holographic duals are given by duality
cascades of gauge theories on D3-branes at the singular geometries.
Generically, steps due to jumps in magnitudes of anomalous mass dimensions appear. Each duality cycle contains a number of steps (the KS throat has only one step in each cycle).
Such steps can introduce sharp features in the cosmic microwave
background radiation. Such a step may have been observed already
\cite{Hinshaw:2006ia, Spergel:2006hy}.
In general, a $D3$ brane encounters many steps as it moves inside a throat.
If the steps are close enough, as expected, the power spectrum may be changed little, but the impact on the non-Gaussianity can be large.
The duality cascade behavior of the warped geometry can therefore have
distinct observable stringy signatures in the cosmic microwave background radiation.
The possibility of detecting and measuring the duality cascade is a strong
enough motivation to study the throat more carefully. Although both Seiberg duality and gauge/gravity
duality are strongly believed to be true, neither has been mathematically proven;
so a cosmological test is highly desirable. This will also provide strong evidence for string theory.
 
It was proposed that one can easily envision a multi-warped throat, that is,
 a throat inside another throat (i.e., the IR of one throat
is matched to the UV of another throat) \cite{Franco:2005fd}. Here we
expect a kink in the warp factor, but not a step. Again, we expect
this to generate some non-Gaussianity.

Here we discuss the generic features introduced by the presence of
such steps in the warped geometry of a throat. We consider the
presence of steps in both slow-roll and DBI inflation. To be general, we shall allow
the steps to go either up or down and we consider both cases. Their properties can be
quite different in these scenarios. However, some features can be
quite independent of details. For example, as an illustration, let us assume the feature
in WMAP CMBR data at $l\simeq 20$ (with width about $\delta l \simeq
10$) is due to such a step. If we further assume that the deviation around
$l \simeq 2$ is at least partly due to another step (i.e., not all due
to cosmic variance), then we can predict that its width $\delta l
\simeq 2$. Furthermore, we would then expect additional steps at $l \simeq 200$
(with width $\delta l \simeq 100$) and $l \simeq 2000$ (with width
$\delta l \simeq 1000$) and so on. However, the sizes of the
steps are more model dependent, though they tend to vary monotonically.
If there is no step at $l \sim 2$, then the next step is expected
at some $l > 200$.

 As illustrations, we consider two specific scenarios:
\begin{enumerate}
\item Its properties in slow-roll brane inflation. Effects of steps on
power spectrum in CMBR in slow-roll inflation have already been
studied in inflationary models \cite{Adams:1997de,Leach:2000yw,Leach:2001zf,Hunt:2004vt}
and in details \cite{Adams:2001vc,Peiris:2003ff,Covi:2006ci} in chaotic
inflation. Its effect on the
non-Gaussianity has also been studied in details in
Ref.~\cite{Chen:2006xj,Komatsu:2003fd}.
Assuming that the $l \sim 20$ feature is due to a step,
Ref.~\cite{Chen:2006xj} shows that the non-Gaussianity in chaotic-like
inflation gives $f_{NL}^{\rm feature} \sim \CO(10)$. Because the
shape and running of this non-Gaussianity is distinctively different
from the cases done in data analysis, whether the PLANCK satellite is going to be able
to test it remains an open question.
In slow-roll brane inflation,
the inflaton, being the position of the brane, is bounded by the size
of the flux compactification volume, and so typically $\phi \ll \Mpl$,
as opposed to large field case for chaotic inflation, where the
inflaton takes values much larger than the Planck mass  ($\phi \ge 15
\Mpl$).
Consequently, the sensitivity of brane inflation to the height and
width of the steps is increased by several orders of magnitude.

However, the appearance of the
warp factor in the potential typically destroys the flatness of the
potential, so some fine-tuning is required.
Here, we shall simply assume slow-roll with steps in the small field case.

\item Properties of steps in the infrared Dirac-Born-Infeld (IR DBI)
model. 
In the UV DBI model, $D$3-branes move down the throat at relativistic
speed \cite{Silverstein:2003hf,Alishahiha:2004eh}.
In terms of known brane inflation scenarios, this model produces too large
a non-Gaussianity to satisfy the current observational bound 
\cite{Chen:2005fe,Baumann:2006cd,Bean:2007hc,Peiris:2007gz} and
inconsistent probe brane backreactions \cite{Bean:2007eh}.
In the IR DBI model,
inflation happens when $D$3-branes are moving relativistically out of a
throat \cite{Chen:2004gc,Chen:2005ad}.
Recent analysis \cite{Bean:2007eh} shows that the IR DBI brane
inflationary scenario fits the CMBR data quite well, comparable to a
good slow-roll inflationary model. However, it has a regional
large running of $n_{s}$: $-0.046 \lesssim d n_{s}/d \ln k \lesssim
-0.01$ at $0.02/\mathrm{Mpc}$ due to a stringy phase transition in
primordial fluctuations. The spectral index has a red tilt $n_{s} \sim
0.94$ at $k \sim 0.02/\mathrm{Mpc}$, and transitions to blue at larger
scales.
It also predicts a large distinct non-Gaussianity due to the DBI
effect, which should be observable by PLANCK. So we see that such IR
DBI models of brane inflation can fit the present cosmological data
and have distinctive stringy predictions different from slow-roll.
In this DBI scenario, steps appear in the warp factor. 
The feature
appearing in the power spectrum has
different properties from that in the slow-roll case.
For example, we can choose the phenomenological parameters that give
reasonable fit to the glitch around $l\sim 20$, but given the extra
parameters that are added, the improvement in $\chi^2$ is not
significant. We will explain why, in general,
the glitches caused by the warp factor steps in IR
DBI model is much smaller than the glitches in slow-roll inflation, while
the associated non-Gaussianity signal can be very large.
In this model, the
non-Gaussianity should come from a combination of DBI relativistic
roll effect together with the step effect, which can be large and is
very distinctive.
\end{enumerate}

As we shall see, some of the predictions are rather sensitive to the details, while others are robust, for example, that there are a number of steps, with their spacing $\Delta\ln \phi \propto \Delta\phi \propto \Delta\ln l$ roughly equal.

 In the IR DBI model with steps, CMBR data can provide the following data points:
$n_{s}$,  $d n_{s}/d \ln k$, and DBI non-Gaussianity, which may be crudely quantified into four data points
(the magnitude as measured by $f_{NL}$, two data points from the shape
of the bi-spectrum, and another for its running\footnote{This
accounting is clearly an oversimplification, since the bi-spectrum is
a function. Furthermore, the 4 point-correlation (tri-spectrum) has
also been calculated \cite{Huang:2006eh}, which may be measured and
tested.}). So the IR DBI scenario alone may confront CMBR via six data
points. Comparing to the slow-roll case, which so far confronts CMBR
via only $n_{s}$, the test of IR DBI is clearly much more
restrictive. Note that the number of microscopic parameters of the
model is roughly five (depending on how one counts) in both brane
inflationary scenarios. Now, let us include the steps. Each step will
provide three data points: its position, height and width, in addition to
the non-Gaussianity associated with the step. The number of
microscopic parameters of the model will increase by at least three. If we
see more than one step in the CMBR, the test of the underlying
superstring theory will be very stringent. In fact, we can learn a lot
about the flux compactification of our universe in superstring
theory. Of course, one may add the cosmic string prediction into the
list to further test the scenarios.

We should mention that, broadly speaking, there are two classes of
inflationary scenarios in string theory, depending on whether the
inflaton is an open string mode (brane inflation) or a closed string
mode (moduli inflation)\footnote{In brane inflation, inflatons are
positions of branes, which are vacuum expectations of light open
string modes. However, the inflaton potential can still be described
by closed string interactions, perturbative, collective and/or
otherwise.}. Moduli inflation can reproduce usual slow-roll inflation,
but so far no distinct stringy signature has been identified.
Obviously it is important to continue the search for distinct stringy
signatures in both brane inflation and moduli inflation.
 
This paper is organized as follows. Section \ref{Sec:Model} reviews
the setup of the geometry and the model. It also reviews $D$3-brane
potential where the step appears as a small correction. Section
\ref{Sec:Pk} presents general formalism on the power spectrum and
bispectrum due
to a sharp step. Section \ref{Sec:srstep} discusses the step in the
slow-roll brane inflation. Section \ref{Sec:irstep} considers a step
in the IR DBI model. We perform qualitative analyses of the step
features in IR DBI inflation followed by numerical analyses. The
comparison to WMAP data around $l\sim 20$ is also performed. 
To be more general, Sec.~\ref{Sec:Pk}, \ref{Sec:srstep} and
\ref{Sec:irstep} are written in a way that the main analyses apply to
general sharp features, such as steps. The duality cascade steps
reviewed in Sec.~\ref{Sec:Model} is used as an example to give a more
restrictive set of predictions. As we emphasized, some details of the
steps such as the width, depth and forms may vary from example to
example.
In Section \ref{Sec:summary} we make some remarks on some broader
aspects of brane inflation. A number
of details are contained in the appendices.

\section{The Model}
\label{Sec:Model}
First it is useful to define a basis of coordinates and a metric.
The ten dimensional metric which describes the throat is that of $AdS_5\times
X_{5}$, where $X_{5}$ has the $T^{1,1}$ geometry in the UV region. Including the expansion of the universe, the metric has the form
\begin{equation}
ds^2=h^{2}(r)(-dt^{2} + a(t)^{2} d{\mathbf x}^{2}) + h^{-2}(r)(dr^2+r^2
ds_{X_{5}}^2),\label{10dmetric}
\end{equation}
Far away from the bottom of the throat,  $ds_{X_{5}}=ds_{T^{1,1}}^2$ is the metric that describes a base of the conifold $T^{1,1}$ which is an $S^3$ fibered over $S^2$.
Here $h(r)$ is the warp factor, that is, a generic mass $m \rightarrow m h(r)$ in the presence of $h(r)$.
 
The inflaton $\phi$ is related to the position of $n_{B}$ 4-dimensional space-time filling $D$3-branes. 
In IR DBI inflation, they are moving out of the $B$ throat into the bulk and then falling into the $A$ throat. 
Inflation takes place while the branes are moving out of the $B$
throat. In UV DBI inflation or the KKLMMT scenario, inflation takes
place when they are moving down the $A$ throat. 
Specifically, we use a single coordinate to describe
collectively the motion of the branes,
\begin{equation}
\phi \equiv \sqrt{n_{B}T_3}r ~.
\end{equation}
The edge of the throat is identified with $\phi_R \equiv \sqrt{n_B T_3}R$, with
\begin{equation}
R^4= \frac{27}{4}\pi g_s N \ap^2, \quad N = KM ~.
\end{equation}
 
In regions where several kinds of 
backreactions from the 4-dimensional expanding background
\cite{Chen:2005ad,Chen:2006ni},
the stringy effects \cite{Chen:2005fe,Chen:2005ad} and the probe-branes
\cite{Silverstein:2003hf,Chen:2004hua} can be ignored,
the following DBI-CS action describes the radial motion of the $D3$ branes,
\begin{equation}
S = \int d^4x \sqrt{-g} \left[ - e^{-\Phi} T(\phi)
\sqrt{1 - \frac{g^{\mu\nu} \partial_\mu\phi \partial_\nu\phi}{T(\phi)}} + T(\phi) - V(\phi) \right] ~.
\label{dbi_cs}
\end{equation}
The warped $D3$ brane tension $T(\phi)$ and the inflaton potential is given by
\begin{eqnarray}
\label{warpedT}
T(\phi) &=& n_{B}T_{3} h^{4}(\phi) ~, \\
V(\phi) &=&  \frac{\beta}{2} H^{2} \phi^{2} + V_{D \bar D}(\phi) ~, \\
V_{D \bar D}(\phi) &=& V_{0}\left(1-\frac{n_{B}V_{0}}{4\pi^{2}v}\frac{1}{(\phi-\phi_{A})^4}\right) ~.
\end{eqnarray}
Here $v$ is the volume ratio; one may take $v =16/27$.
For our purpose here, the Coulomb term is negligible so $V_{D \bar
  D}(\phi) =V_{0}= 
 2n_A T_{3} h^{4}(\phi_{A})$ is the effective vacuum energy. We expect
 $|\beta| \sim 1$. Positive $\beta \gg 1$ for UV DBI model while
 negative $\beta \sim -1$ for IR DBI model. Slow-roll inflation
 requires $|\beta| \ll 1$.
 
It is convenient to introduce the inflationary parameter $\epsilon \equiv -\dot{H}/H^2$, so that
\begin{equation}
\frac{\ddot a}{a} = H^{2}(1 - \epsilon) ~.
\end{equation}
The universe is inflating when $\epsilon < 1$. In all the scenarios
that we consider here, $\epsilon$ never grows to 1, so inflation ends
by $D\bar{D}$ annihilation.
 
\begin{figure}[th]
\begin{center}
\includegraphics[width=10cm]{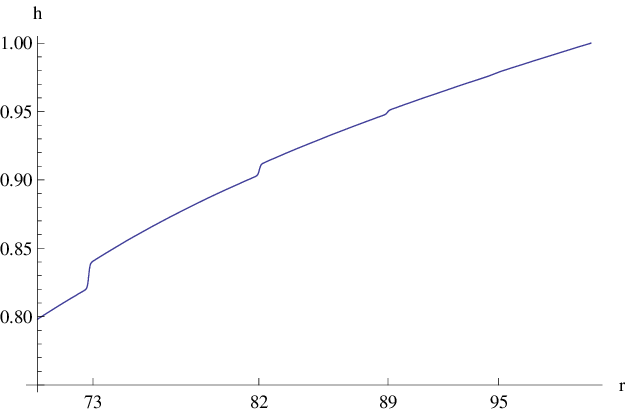}
\end{center}
\caption{The warp factor $h(r)$ in the KS throat, including the steps. Here $r$ is measured in $\sqrt{\ap}$ with $R_B \simeq 100$, $g_s=2$, M=20$, K=10$ and $N=200$. The width $d=10^{-3}$. In this figure, there are actually 4 steps, located at $r \simeq 73$, $r \simeq 82$,  $r \simeq 89$ and $r \simeq 95$, although the step at $r \simeq 95$ is too small to show up.
Here, the parameters (in particular, a large $g_s M$) are chosen so that at least 3 steps are big enough to show up in the figure. This leads to relatively large corrections to the positions of the steps. Other parameters should be used in more realistic situations and in comparison with data. }
\label{stepsketch}
\end{figure}

For simplicity, let us assume that the B throat is a KS throat.
The warped factor for the $B$ throat around the $p_l^\mathrm{th}$ duality transition (starting from the bottom of the throat) is simplified to (see Appendix \ref{Appwarp}) \cite{Hailu:2006uj}
\begin{equation} \label{step_b}
h^{4}(r) \simeq \frac{r^{4}}{R_B^4} \frac{K}{p_l} (1 + \Delta) ~,
\quad
\Delta = \sum_{p_{i}}^{K}  \frac{3g_{s}M}{16\pi} \frac{1}{ p^3}
\left[1 + \tanh \left(\frac{r-r_{p}}{d_p}\right) \right] ~,
\end{equation}
where $h(r=R_B) \simeq 1$ at the edge of the throat. 
Here $r_p$ is the positions of the steps, the initial $p_{i}
\gg 1$ so that the warped factor formula is approximately good, and
$d_p$ controls the width of the step. 
The steps has a constant separation in $\ln r$, that is,
$\ln r_{p+1} - \ln r_p \simeq 2 \pi/ 3 g_s M$.
As one moves down the throat ($r$ and $p$ 
decreasing), note that the step in the warped factor $h^{4}(r)$ of the
KS throat is going down. See Figure \ref{stepsketch}. This stepping down happens at each Seiberg
duality transition. Together, they form a cascade
\cite{Strassler:2005qs}. For large $K$, one encounters $K-1$ steps as
one approaches the infrared. Note that we are ignoring smooth corrections to the shape of the warp factor even though they may be larger than the step size. This is a reasonable approximation since it is the sharp features that will show up as distinctive features in the CMBR. 
 
Steps are a generic feature: consider another throat whose gauge dual has $n_{G}$ gauge factors,
with appropriate bi-fundamentals \cite{Franco:2005fd}. As the gauge
coupling of the gauge factor with the fastest running towards large
coupling in the infrared (decreasing $r$) gets strong, Seiberg duality
transition applies.
This happens for each gauge group factor sequentially until we reach
the same structure as the original gauge model. Typically, it takes
$n_{G}/n_{s}$ number of transitions to complete this cycle, where
$n_{s}$ is a factor in $n_{G}$ (the KS throat has $n_{G}= n_{s}=2$ so
each cycle has only 1 transition). Because of the jump in the
corresponding anomalous mass dimension, steps in the warp factor are
very generic.
It is also likely that for another throat with a different geometry,
some steps may show up with an opposite sign. To be general, we shall
discuss each of these scenarios.
If the throat has relatively few steps, or if the steps are well
separated, the impact of individual step on CMBR may be observable.
Otherwise, the steps may be too close for them to show up in the power
spectrum.
 
In the original KS throat solution, they take the approximation
where the dilaton is constant. Here the dilaton factor, $e^{-\Phi}$,
runs in general and is $\phi$ dependent. However, for the IR DBI
inflation model we discuss in this paper, most of the DBI e-folds are
generated at the tip of the throat where $HR_B^2/N_e < r < HR_B^2$. At
the same time, the brane moves across roughly $g_s M$ steps
\cite{Bean:2007eh}. The running of the dilaton can be ignored if $g_s M
\ll p_l$, which is most easily satisfied with a small
$g_s$. Furthermore,
the WMAP data only covers a few e-folds, which corresponds
to $3g_s M/N_e$ steps, so we only need $p_l \gtrsim g_s M$ 
to safely ignore the
dilaton modification to the kinetic term. In this paper, when we
discuss the IR DBI inflation, we always absorb $e^\Phi$ into $g_s$ as
a constant, so it will never appear explicitly in our analysis. 

The energy density and pressure from the DBI action are given by
\begin{eqnarray} \label{dbi_rho_p}
\rho = V(\phi) + T(\phi)(c_s^{-1} - 1), \quad p = -V(\phi) + T(\phi)(1-c_s),
\end{eqnarray}
where the sound speed is defined as the inverse of the Lorentz factor $\gamma$,
\begin{equation}
c_{s}=\gamma^{-1} = \sqrt{1- \dot{\phi}^2/T} ~.
\end{equation}
The background equation of motion is given by
\begin{eqnarray}
& &V(\phi) + T(\phi)(c_s^{-1} - 1) = 3H^2 ~,\\
& &\ddot\phi - \frac{3}{2}\frac{T'(\phi)}{T(\phi)}{\dot\phi}^2 + 3H c_s^2\dot\phi + T'(\phi) + c_s^3[V'(\phi) - T'(\phi)] = 0 ~. \label{dbi_eom}
\end{eqnarray}

On the other hand, the running dilaton is more relevant for the UV DBI
or the KKLMMT scenario, because here inflationary e-folds are generated
by the $D3$ brane moving all the way from the UV to the IR end of the
throat. For the KKLMMT scenario, we expand the DBI action in the
$\dot{\phi}^2 \ll T(\phi)$ limit, and get the slow-roll action,
\begin{eqnarray}\label{action_sr}
S_{SR} = \int d^4x\, \sqrt{-g} \left[ \frac{1}{2}e^{-\Phi} \dot{\phi}^2 - T(\phi)(e^{-\Phi} - 1) - V(\phi) \right] ~.
\end{eqnarray}
We see explicitly that the effective inflaton potential,
$V_{\mathrm{eff}}(\phi) = T(\phi)(e^{-\Phi} - 1) + V(\phi)$, now
exhibits the step feature from $T(\phi)$. In the original KS solution,
$e^{-\Phi} = 1$, and the steps in the warp factor do not affect the
inflaton potential. The term $T(\phi)(e^{-\Phi} - 1)$ is
generically bad for slow-roll inflation, however. Since $T(\phi) \sim \phi^4$,
we have effectively introduced a quartic term in the potential. It
is well known that a $\lambda\phi^4$-type potential requires a
trans-Planckian $\phi$ to generate inflation. However, in the KS
throat (and other GKP type compactifications \cite{Giddings:2001yu}), trans-Planckian $\phi$
is impossible \cite{Chen:2006hs, Baumann:2006cd} and we have a strong
bound on the largest inflaton field value,
\begin{eqnarray}
\frac{\Delta\phi}{\Mpl} \lesssim \frac{1}{\sqrt{KM}} \ll 1 ~.
\end{eqnarray}
In order to get slow-roll inflation, the dominant term in the potential has to be the uplifting term from $\bar{D}3$ at the bottom of the throat, so we require
\begin{equation}
V_0 \gg T(\phi)(e^{-\Phi} - 1) ~.
\end{equation}
Since $V_0 \sim h^4(\phi_A) \ll T(\phi) \sim h^4(\phi)$, the above relation is true only when $
e^{-\Phi} \approx 1$.
Note that if $V_0$ dominates the inflaton potential, the step height in the inflaton potential is not given by $\Delta$ as defined in Eq.(\ref{step_b}), but is suppressed by a factor $(e^{-\Phi} - 1)\phi^4/\phi_A^4 \ll 1$, i.e.
\begin{equation}
\frac{\Delta V}{V} = (e^{-\Phi} - 1)\frac{\phi^4}{\phi_A^4} \Delta ~.
\end{equation}
So far, we have argued that even if the dilaton runs generically, slow-roll inflation can only occur within the region where
$e^{-\Phi}$ stays close to 1. Thus when we discuss slow-roll scenario
in this paper, we will ignore the $e^{-\Phi}$ modification to the
kinetic term, and set $e^{-\Phi} \dot{\phi}^2 \approx \dot{\phi}^2$,
but we keep the term $T(\phi)(e^{-\Phi} -1)$ in the
potential. This approximation should be pretty good, since $e^{-\Phi}$
only exhibits mild features like kinks, while $T(\phi)$ has sharp
features like steps. When both of them are present, the sharp features
in $T(\phi)$ definitely dominate.

\section{The Power Spectrum and Bispectrum in General}
\label{Sec:Pk}
The gauge invariant scalar perturbation $\zeta(\tau,\bk)$ can be
decomposed into mode functions
\begin{eqnarray}
\zeta(\tau,\textbf{k})=u(\tau,\textbf{k})a(\textbf{k})
+u^*(\tau,-\textbf{k})a^{\dagger}(-\textbf{k}) ~.
\end{eqnarray}
The mode function $u_\bk(\tau)$ is given by the equation of motion
\begin{equation}
v_k'' + \left( k^2 c_{s}^{2} - \frac{z''}{z} \right) v_k =0 ~,
\label{quadeom}
\end{equation}
where
\begin{equation}
v_k\equiv z u_k ~, \quad z \equiv a \sqrt{2\epsilon}/c_{s} ~,
\label{vdef}
\end{equation}
and prime denotes derivative with respect to the conformal time $\tau$
(defined as $dt \equiv ad\tau$). 
In the absence of any features in either potential or warp factor, 
the power spectrum is given by
\bea
P_{\calR}(k) &\equiv& \frac{k^3}{2\pi^2} |u_\bk|^2
\label{powerspec} \\
&=& \frac{H^2}{8\pi^2\Mpl^2 \epsilon c_s} ~,
\eea
where $u_\bk$ is $u(\tau, \bk)$ evaluated after each mode crosses the horizon.
 
To study the effects of features, we define the inflationary parameters as below
\be \label{eps_def}
\epsilon \equiv -\frac{\dot H}{H^2} ~, \quad
\tilde{\eta} \equiv \frac{\dot \epsilon}{H\epsilon} ~, \quad
s \equiv \frac{\dot c_s}{H c_s} ~.
\ee
We can express $z''/z$ exactly as,
\begin{eqnarray}
\frac{z''}{z} = 2a^2H^2 \left( 1- \frac{\epsilon}{2} + \frac{3{\tilde{\eta}}}{4}
-\frac{3s}{2} -\frac{\epsilon \tilde{\eta}}{4} +\frac{\epsilon s}{2}
+\frac{{\tilde{\eta}}^2}{8} -\frac{\tilde{\eta} s}{2} +\frac{s^2}{2}
+\frac{\dot{\tilde{\eta}}}{4H} -\frac{\dot s}{2H} \right) ~.
\label{z''/z}
\end{eqnarray}
The
$z''/z$ encodes all the information from  the inflationary background,
and determines the evolution of $u(\tau,\bk)$. In the absence
of sharp features, $\epsilon$, $\eta$ and $s$ remains much smaller
than 1, so $z''/z \sim 2a^2H^2$. However, a sharp feature in the
inflation potential or the warp factor will induce a sharp local
change in $\epsilon$, $\tilde \eta$ and $s$, and $z''/z$ has a nontrivial
behavior deviating strongly from $2a^2H^2$ around the feature. In this
paper, we have analyzed two cases. First, the slow roll brane
inflation with a step feature in the inflaton potential $V(\phi)$. Due
to the small field nature of brane inflation, $\epsilon$ is
negligible, which greatly simplifies $z''/z$ to (See Appendix
\ref{AppRelax})
\begin{equation}
\frac{z''}{z} \approx 2a^2H^2 \left(1 -\frac{V''(\phi)}{2H^2} \right) 
= 2a^2 H^2 \left(1-\frac{3}{2}\frac{V''}{V} \right)
~.
\label{z''/zSR}
\end{equation}
Second, the IR DBI brane inflation scenario with a sharp step
appearing in the warp factor. In Section \ref{irdbi_pk}, we show that
the sharp change of the sound speed is the major contribution to
$z''/z$, and we have
\begin{equation}
\frac{z''}{z} \approx 2 a^2 H^2 \left(1-\frac{\dot{s}}{2H} \right)
=2a^2 H^2 \left(1 - c_s\epsilon \frac{T''}{T} \right)
~,
\label{z''/zDBI}
\end{equation}
where $T$ is the warped brane tension defined in Eq.(\ref{warpedT}).
To avoid confusion, we emphasize that the primes on $z$ or $u_k$
(functions of time) denote the derivative with respect to the conformal
time $\tau$, while the primes on $T$ or $V$ (functions of $\phi$)
denote the derivatives with respect to the
field $\phi$.
  
Among all the terms in the exact cubic action for the
general single field inflation \cite{Chen:2006nt},
the 3pt for sharp features receives dominant contribution from the
term proportional to $\frac{d}{d\tau} (\frac{\tileta}{c_s^2})$ in most
interesting cases,
\begin{eqnarray}
\langle\zeta^3\rangle
&=& i \left( \prod_i u_{k_i}(\tau_{end}) \right)
\int_{-\infty}^{\tau_{end}} d\tau a^2 
\frac{\epsilon}{c_s^2} \frac{d}{d\tau} \left(\frac{\tileta}{c_s^2}\right)
\left( u_{k_1}^*(\tau) u_{k_2}^*(\tau)
 \frac{d}{d\tau} u_{k_3}^*(\tau) + \mathrm{perm} \right) 
\nonumber \\
&\times& (2\pi)^3 \delta^3 (\sum_i
\bk_i ) + \mathrm{c.c.} ~.
\label{3ptGeneral}
\end{eqnarray}
The reason is a generalization of that in Ref.~\cite{Chen:2006xj}. As
we will show, $\epsilon$ and $c_s$ do not change much due to
approximate energy conservation relations; but they change in a very
short period of time, boosting $\dot \epsilon$ or $\dot c_s$ by
several orders of magnitude.

\section{Steps in Slow-Roll Brane Inflation}
\label{Sec:srstep}
In the slow-roll scenario, let us assume that inflation takes place as the brane moves down the throat. Since the dilaton runs, the slow-roll potential is given by
\begin{equation}
V_{SR} (\phi) = V(\phi) +T(\phi)(e^{-\Phi (\phi)}-1) ~.
\end{equation}
Here, the second term in the potential does not vanish, so the step in $T(\phi)$ shows up in the potential. Again, we may consider both step up and step down cases.
 
Before we proceed, we mention two other possibilities that will not be
studied in more detail in this paper.
Suppose the warp factor steps down as $r$ increases. As the inflaton moves down the throat, it encounters such steps. If the height of any particular step is too big, then the inflaton cannot overcome it so its classical motion towards the bottom of the throat is stopped. Let us say this happens at $r=r_{p}$. It can still move along the $S^{3}\times S^{2}$ angular directions. In a realistic model, the
$S^{3}\times S^{2}$ symmetry is slightly broken so the inflaton would
move along the steepest angular direction, until it reaches the lowest
point at this radius $r_{p}$. If the step there is lessened so that
the inflaton can roll over it, the inflaton can continue to roll down
along $r$ until it reaches the next step. This process can repeat some
number of times. As a result, the inflaton path is substantially
longer than just moving down along $r$. Effectively, this may
substantially extend the field range and enough e-folds of inflation
is much more likely. 
 
Another possibility is that the brane can tunnel over the step
barrier. Recent analysis of the tunneling of branes suggests that,
under the right condition, the brane may tunnel very easily across a
potential barrier that is not particularly small
\cite{Brown:2007ce}. If this is the case, then the brane may roll
along some angular direction until it reaches such a position and
tunnel right over the potential barrier there. It can then roll down
along $r$ until it reaches the next step. Again, this process can
repeat any number of times, as a result, the inflaton path is again
substantially longer than just moving down along $r$, allowing the
generation of many e-folds.
 
In the following, we consider the scenarios where the inflaton can roll over many steps in the potential.

\subsection{The Position of the Steps}
\label{PositionofSteps}
 The duality cascade predicts more than one step,
so it is interesting to ask where the steps are. Assume there is a step at
the $l \simeq 20$ in the CMBR, which may be checked with a non-Gaussianity
measurement. Let us further assume that there is another step at $l
\simeq 2$, (the significance of the signal at $l=2$ in power spectrum is much smaller than
that in $l\simeq 20$ due to the large cosmic variance, so we
assume this mainly for the purpose of illustration).
We now show that the location of all the steps in the duality cascade
can be determined by given the locations of two such adjacent steps. 

Using $l \sim 10^{4} (k/\mathrm{Mpc}^{-1})$, we find
\begin{eqnarray} \label{dne}
-dN_{e} \simeq d \ln k \simeq d\ln l \simeq H dt \simeq \frac{H}{\dot
 \phi} d\phi ~,
\end{eqnarray}
Since both $H$ and $\dot \phi$ are slowly varying during inflation,
we have
\bea
d\ln l \propto d\phi ~.
\label{lphi}
\eea
Suppose $\phi$ is decreasing (going down a throat), also suppose that the step at $l=2$ is at $\phi_{m_{0}}$ and that at $l=20$ is at $\phi_{m_{0}+1}$,
we have
\be
\frac{\phi_{m_{0}}}{\phi_{m_{0}+1}}  \simeq \frac{\phi_{m_{0}+1}}{\phi_{m_{0}+2}} \simeq ... \simeq e^{2 \pi/3 g_{s}M}.
\ee
For large $g_{s}M$, this ratio is close to unity, $e^{2 \pi/3 g_{s}M} \simeq 1 + \delta$, so
$\phi_{m_{0}} - \phi_{m_{0}+1} \simeq \phi_{m_{0}+1} - \phi_{m_{0}+2} \simeq \phi_{m_{0}+1}\delta$.
Due to (\ref{lphi}),
equal spacing in $\phi$ implies equal spacing in $\ln l$.
So we find that the next two steps are at around $l \simeq 200$ and $l \simeq 2000$ respectively.
In addition, the effect of the step at multiple moment $l$
should span over $\Delta l$ multiple moments with $\Delta l
\propto l$.

\subsection{The Power Spectrum}
\label{SecPowerSR}
In the slow-roll scenario without features, we have the attractor
solution, in units of $M_{P}$,
\be
H^{2}=V(\phi)/3, \quad  3H {\dot \phi} =-V'(\phi) ~,
\ee
where $\prime$ is derivative with respect to the inflaton $\phi (t)$
and dot is derivative with respect to time. Here,
\be
\epsilon = -\frac{\dot H}{H^{2}} = 2\left(\frac{H'}{H} \right)^{2} = \frac{1}{2}\left(\frac{V'}{V} \right)^{2}
=\epsilon_{SR}.
\ee
It is convenient to introduce another inflationary parameter
\be
\tilde{\eta} = \dot \epsilon/H\epsilon = -2 \eta_{SR} + 4 \epsilon_{SR}
\ee
where the usual slow-roll version is given by
\be
\eta_{SR}= \frac{V''}{V}.
\ee
This yields
$n_{s} -1 = - \tilde{\eta} -2 \epsilon = 2 \eta_{SR} - 6 \epsilon_{SR}$.
We emphasize that the above relations between $\epsilon$,
$\tileta$ and $\epsilon_{SR}$, $\eta_{SR}$ only hold for the
attractor solution in absence of sharp features. We will use the
parameters $\epsilon$ and $\tileta$ in our analyses for the sharp
feature case.

To see the full details of the effect of a step in potential,
numerical calculation is necessary. However, the qualitative behavior
can be estimated as follows.
 The step in the potential
is typically characterized by two numbers: the
depth which we describe as the ratio $\Delta V/V \approx 2 c$, and
the width $\Delta \phi =2d$ in unit of Planck mass.
We can divide the motion of inflaton into two parts: acceleration and
relaxation.
First, the inflaton, originally moving in
its attractor solution, momentarily gets accelerated by the step. The
potential energy $\Delta V$ gets converted to kinetic energy. Then
after the inflaton moves across the step, it starts to relax back to
its attractor solution under $\ddot \phi +3H\dot \phi \approx 0$,
where the Hubble friction term dominates over the potential.

We first look at the power spectrum.
The inflaton velocity in the original attractor solution is given by
$|\dot\phi_{attr}| \approx |V'|/3H \approx \sqrt{\epsilon V/3}$.
After the acceleration of the step, $\dot \phi \approx
\sqrt{V(3c+\epsilon/3)}$. (It turns out that
the best fit data give comparable $3c$ and $\epsilon/3$, so
for the purpose of order-of-magnitude estimate we later will also
approximate  $\dot \phi$ to be $\sqrt{3cV}$.)
A useful formula to understand the dip of the glitch in the
power spectrum caused by the step is $P_\calR = H^2/(4\pi^2\dot
\phi^2)$. The acceleration of inflaton does not change the $H$ so it
reduces the $P_\calR$. The ratio of $P_\calR$ between the initial value and
the value at the first dip is related to the corresponding ratio of $\dot
\phi$, which is $\sqrt{1+9c/\epsilon}$. For the large field quadratic
inflation,
Ref.~\cite{Peiris:2003ff,Covi:2006ci} 
give the best-fit data, $c=0.0018$,
$\epsilon=0.009$. As we will see later, the values of $c$ and
$\epsilon$ will be rather model-dependent. However the discussion in
this paragraph shows that their ratio should be fixed for a specific
observed feature,
\be
\frac{c}{\epsilon} \approx 0.2 ~.
\label{cepsilon}
\ee

\begin{figure}[th]
\begin{center}
\includegraphics[width=15cm]{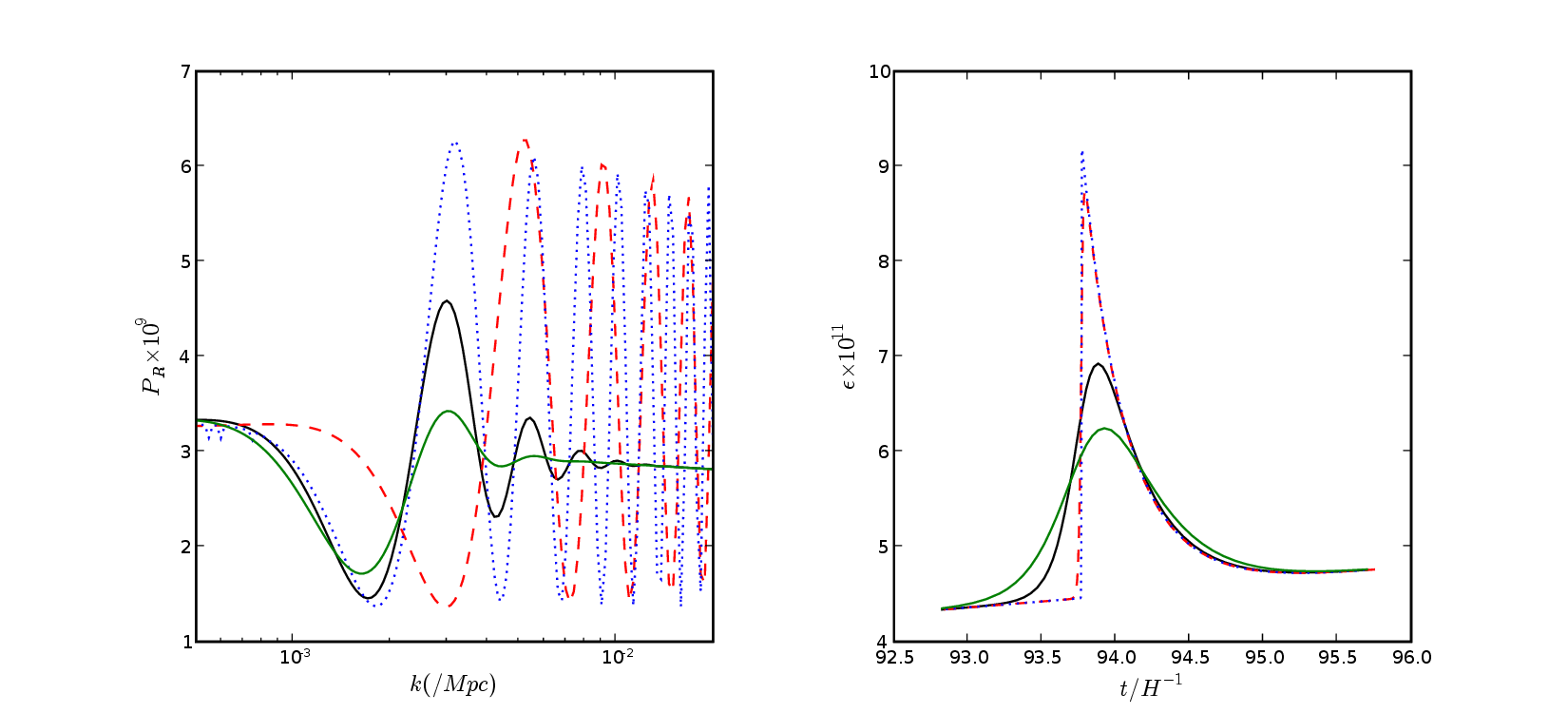}
\end{center}
\caption{\small In the small field case, we show how the step in the
inflaton potential changes the power spectrum. We show the power
spectrum in the left panel and the behavior of $\epsilon$ around the
step in the right panel. For illustration, we use the KKLMMT scenario
with only the Coulomb potential. We choose the background flux
$N=2000$, so $\Delta\phi/\Mpl \lesssim 0.01$. The inflation scale $V_0
\sim 10^{-17}\Mpl^4$. $\epsilon$ is tiny, typically $\epsilon \sim
10^{-11}$ as shown in the right panel. We introduce a step with $c =
8\times10^{-12}$ and calculate the power spectrum for four different
values of $d$: (1) $d=3\times 10^{-6} \Mpl$, green solid line.
(2) $d=1.7\times 10^{-6} \Mpl$, black solid line. (3) $d=1.7\times
10^{-7}\Mpl$, red dashed line. (4) $d=1.7\times
10^{-8}\Mpl$, blue dotted line. Note that the dip in
the power spectrum depends weakly on $d$. In case (1) and (2), $\Delta\epsilon$ 
has not saturated the bound Eq.~\protect\ref{deltaepsilon}, so decreasing $d$ will enhance the bump
in $P_\calR$ significantly. In case (3) and (4), where $\Delta\epsilon$ is maximized, the bump in 
$P_\calR$ does not depend sensitively on $d$, but the range of the oscillations in
$k$-space does. The black solid line is close to the best fit power 
spectrum given in Ref.~\protect\cite{Peiris:2003ff,Covi:2006ci}.}
\label{sr_Pk}
\end{figure}

Although the above discussion indicates that the depth of the dip is relatively 
independent of $d$, the peak of the first bump is.
When the inflaton moves across the step, energy conservation requires that $\frac{1}{2} \Delta \dot\phi^2 \le \Delta V$, with the bound saturated when the Hubble friction is negligible. This will result in an upper bound in the change of $\epsilon$ 
\begin{equation}
\Delta \epsilon \approx \Delta V/H^2 \lesssim 5c ~.
\label{deltaepsilon}
\end{equation}
We expect that the effect of decreasing $d$ will first enhance the
bump, since smaller $d$ leads to more deviation from the attractor
(larger $\Delta\epsilon$), and will enhance the bumps in $P_\calR$
during the relaxation period. However, as long as $\Delta\epsilon$
saturates the upper bound (when $d$ is small enough to ignore the
Hubble friction), further decreasing $d$ will not affect the bumps in
$P_\calR$ any more, since the relaxation period essentially starts
with the same $\dot\phi$ no matter how small $d$ is. So we expect the
bump to depend sensitively on $d$ if moving across the step takes
$\calO(1)$ e-folds, and it becomes relatively insensitive to $d$ when
the step is so sharp that moving across it takes only $\ll 1$
e-folds. In the latter situation, reducing $d$ will only increase the
extension of the oscillations in $P_R$,
\bea
\Delta k \sim \sqrt{\frac{z''}{z}} \sim k_0 \frac{\sqrt{c}}{d} ~,
\label{kextSR}
\eea
where $k_0$ is the starting point of the feature, and (\ref{z''/zSR}) is used.

Ref.~\cite{Peiris:2003ff,Covi:2006ci} fit the $l \sim 20 $
feature in the WMAP data introducing a step feature in the large field slow-roll
model and gives the best fit power spectrum. Here we have
numerically reproduced a similar power spectrum for the small field
case. We are not performing a complete data analysis to find the best
fit model here, our major emphasis is to show how the width $d$ comes
into play. Fig.~\ref{sr_Pk} shows the power spectrum and $\epsilon$
for four different values of $d$. The black solid line represents the
power spectrum close to the best fit model given in
Ref.~\cite{Peiris:2003ff,Covi:2006ci}, with $c=8\times 10^{-12}$, we
find that $d=1.7\times 10^{-6} \Mpl$.

\subsection{Non-Gaussianities}
Now let us look at the bispectrum.
For slow-roll inflation, the 3-point function of the scalar perturbation
$\zeta(\tau_{end},\bk)$ includes terms proportional to $\epsilon^2$,
$\epsilon^3$, and $\epsilon \tilde{\eta}'$ \cite{Maldacena:2002vr}. 
For the slow-roll potential without
any features, the first terms dominate. Comparing to WMAP's ansatz,
it gives the non-Gaussianities estimator $f_{NL} = \CO(\epsilon)$.
 
In the presence of sharp features, from (\ref{deltaepsilon}) we see
that $\epsilon$ still remains small, $\lesssim \CO(0.01)$. But
$\tilde{\eta}'$ can be much larger.
Therefore, the leading three-point correlation function is given by the term
proportional to $\epsilon \tilde{\eta}'$ \cite{Chen:2006xj},
\begin{eqnarray}
&&\langle\zeta(\tau_{end},\mathbf{k}_1)\zeta(\tau_{end},\mathbf{k}_2)\zeta(\tau_{end},\mathbf{k}_3)\rangle
  \nonumber \\
&=& i \left( \prod_i u_{k_i}(\tau_{end}) \right)
  \int_{-\infty}^{\tau_{end}} d\tau a^2 \epsilon {\tileta}' \Big(
  u_{k_1}^*(\tau) u_{k_2}^*(\tau) \frac{d}{d\tau} u_{k_3}^*(\tau) +
  \mathrm{perm} \Big) \nonumber \\
&\times& (2\pi)^3 \delta^3 (\sum_i \bk_i) + \mathrm{c.c.}~,
\label{term1}
\end{eqnarray}
where the ``perm'' stands for two other terms that are symmetric under
permutations of the indices 1, 2 and 3.
The details of such an integration are quite complicated. Nonetheless
we can estimate the order of magnitude of the
non-Gaussianity estimator $f_{NL}$ by comparing it to the slow-roll case.
The most important difference is that here we
replace a factor of $\epsilon$ by $\tilde{\eta}'$, and also $\tilde{\eta}'$ only gets
large momentarily. Hence we estimate 
$f_{NL}^{\rm feature} = \CO(\tilde{\eta}'\Delta \tau) =
\CO(\Delta \tilde \eta)$,
where $\Delta \tau$
is the conformal time that the inflaton spends crossing the step.

To estimate the level of these non-Gaussianities,
we now give a qualitative estimate for $\tilde{\eta}'$ \cite{Chen:2006xj}.
During the acceleration period, the
$\epsilon$ is increased by
\be
\Delta \epsilon \approx \Delta V/H^2 \approx 5c ~.
\ee
The duration of this period is
\be
\Delta t_{accel} \approx \Delta \phi/\dot
\phi \approx d/\sqrt{c V} ~,
\ee
where we used  $\dot \phi$ estimated in Sec.~\ref{SecPowerSR}.
These can be further used to
estimate
\be
\Delta \tilde \eta \approx \tilde{\eta} = \frac{\dot \epsilon}
{ H\epsilon} \approx \frac{7 c^{3/2}}{d\epsilon} ~,
\label{eta}
\ee
The time scale for the relaxation period is of order $H^{-1}$, during
which $\tilde{\eta}$ is $\CO(1)$ and $\dot
{\tilde\eta}$ are of order $\CO(H)$.
 
We sum over the contributions from both the acceleration and
relaxation periods. The former gives $f_{NL}^{\rm accel} \approx
7c^{3/2}/(d\epsilon)$,  the
latter gives $f_{NL}^{\rm relax} = \CO(1)$. So for most interesting cases
where non-Gaussianities are large enough to be observed,
it can be estimated by the first contribution,
\be
f_{NL}^{\rm feature} \sim \frac{7c^{3/2}}{d\epsilon} ~.
\label{fNLSR}
\ee
We note that this is only a crude order of magnitude estimation on the
amplitude of this non-Gaussianity, since the integration
(\ref{term1}) also involve mode functions $u_k$ which will be
modulated by the presence of the sharp feature, and the shape and
running of such non-Gaussianities are very
important. Details have to be
done numerically, as in \cite{Chen:2006xj}.
Qualitatively since we have argued in Sec.~\ref{SecPowerSR} that for a
specific observed feature the $c/\epsilon$ is hold fixed
model-independently, (\ref{fNLSR}) implies that the value of $d$
is very crucial to the level of the non-Gaussianities.
There are two major constraints on $d$. First, in the power spectrum,
as shown in Fig.~\ref{sr_Pk} the bump in $P_\calR$ may depend sensitively on $d$. 
Second, the range of oscillation in $P_\calR(k)$ is also controlled by $d$. If $d$ is too small,
the oscillation in $P_\calR(k)$ might spread over to the well measured
first acoustic peak in WMAP $C_l$ curve. Numerically, we have found
that $\sqrt{c}/d \sim \calO(1)$, consistent with (\ref{kextSR}), 
so the magnitude of (\ref{fNLSR})
should be close to that in Ref.~\cite{Chen:2006xj}.

It is instructive to split this expression to $f_{NL}^{\rm feature}
\sim 7 c/\epsilon \cdot \sqrt{c}/d$. The first factor is determined by
the amplitude of the glitch from (\ref{cepsilon}), while the second
factor is the extension of the glitch from (\ref{kextSR}). Note both are
in the $k$-space not the CMB multipole $l$-space.
Therefore in principle a sharp feature can also appear
{\rm only} in the 3-point function. This is clear from our estimation (\ref{cepsilon}) 
and (\ref{fNLSR}), where one
can reduce $c/\epsilon$ while increase $c^{3/2}/(d\epsilon)$.

Our analyses so far do not depend on the whether the inflation is
caused by large field or small field, but when it comes to the
quantitative analyses the differences are interesting.
For large field inflation $\epsilon \sim \eta$ while for
small field inflation $\epsilon \ll \eta$.
For example, previous numerical works focus on the large field quadratic
potential, and give the best fit $\Delta V/V \sim c \sim 0.2\epsilon
\sim \CO(10^{-3})$.
Brane inflation is a small field inflation \cite{Chen:2006hs} and
$\epsilon$ is much smaller, $\epsilon \sim \CO(10^{-6})$ or even
$\CO(10^{-12})$. Similarly the field range $d$ (in units
of $\Mpl$) is also
much smaller. So brane inflation is sensitive to very
tiny steps present in the potential. This can be potentially
used as a sensitive probe to the fundamental theory.

\section{Steps in IR DBI Inflation}
\label{Sec:irstep}
The DBI-CS action is Eq.(\ref{dbi_cs}) with $V(\phi)= V_0
- \half m^2 \phi^2 $. The branes start from the tip of the throat and
end at the UV end of the throat $\phi_{\rm end} = R_B \sqrt{n_B T_3} =
\sqrt{\lambda_B}/R_B$.
After that, branes will quickly find antibranes and annihilate.
 
In the relativistic (DBI) regime with $t \ll -H^{-1}$, the attractor solution is
\begin{eqnarray}\label{irdbi_attrac}
\phi \simeq -\frac{\sqrt{\lambda_{B}}}{t} \left(1 - \frac{9}{2 \beta^{2}H^{2} t^{2}}\right) ~, \quad
\gamma \simeq \frac{\beta H}{3} |t| ~.
\end{eqnarray}
This zero mode evolution relies on the background warped geometry, so
it is reliable in regions where the back-reaction deformation from
Hubble expansion is small \cite{Chen:2005ad,Chen:2006ni}. 
This still leaves at least two interesting
regions for primordial fluctuations: the field theoretic region
(corresponding to smaller $N_e$ and shorter scales) and the stringy
region (corresponding to bigger $N_e$ and larger scales). The spectral
index transitions from red to blue \cite{Chen:2005ad,Chen:2005fe}.
Detailed estimates can be found in Ref.~\cite{Bean:2007eh}.

We now consider the effects of sharp features in the warped geometry on this IR DBI model.
The duality cascade will leave sharp features on
the warp factor $T(\phi)$ but not the inflaton potential $V(\phi)$.
Since the WMAP window spans a few e-folds, corresponding to  branes
rolling across $\Delta l \approx 3g_s M/N_e^{\rm DBI}$  duality
cascades in a Klebanov-Strassler throat.
In this paper we will mainly treat the steps as individual sharp
features, namely they are well separated in terms of e-folds and
sufficiently sharp to generate observational signature
individually. For this purpose, we would like to consider the situation where
\be
g_s M \lesssim N_e^{\rm DBI} ~.
\label{gsMcond}
\ee
In the multi-throat brane inflation scenario, the number of inflaton
branes generated in B-throat is roughly determined by the flux number
$M$. Since the number of inflaton branes is constrained to be around
$10^4 \sim 10^5$ \cite{Bean:2007eh}, 
the condition (\ref{gsMcond}) implies a small $g_s \lesssim 10^{-3}$. 

It is also very natural that $g_s$ takes value larger than $10^{-3}$. 
In this case, the branes come across many
steps within a single e-fold. We will discuss this case in
Sec.~\ref{Sec:Res}.

\subsection{The Properties of the Steps}
 
Here we like to make a crude estimate on the positions, fractional
heights and widths of the steps, along the line of
Sec.~\ref{PositionofSteps} but with more detailed numbers.
The input is the position, fractional height and width of the step at
$l\sim 20$ as well as the position of the assumed step at $l \sim
2$.
Let $r=r_{0} \simeq R_{B}$ be the edge of the throat and
$r=r_{b}$ at the bottom of the throat.  They are related by the warp factor $h(\phi_{b})=h_{B}$,
\begin{equation}
r_{b} \sim r_{0}h_{B}\sim r_{0} \exp\left(-\frac{2\pi K}{3g_s
  M}\right) ~.
\end{equation}
We are interested at regions where $r_{b} \ll r \ll r_{0}$.
Thus, given $r_{0}$ or $r_{b}$, we can determine all the other
transition values $r_p$. Here $p$ measures (starting at the bottom) the step in the duality
cascade. In the large $K \gg p \gg 1$ limit, the correction is small, so we treat the steps as  perturbations.
The sharpness of the steps are controlled by the coefficient $d_{p}$.
A step becomes infinitely sharp as $d_{p} \rightarrow 0$. We shall
leave $d_p$ as free
parameters. Note we can introduce similar spreads to the running of the dilaton.
 
The position $\phi_p$ of the steps can be computed. The steps are at
$\phi=\phi_{p}$, where
\begin{equation}
\label{rlrK}
\ln {\frac{\phi_{p}}{\phi_{b}}} \simeq \frac{2 \pi }{3 g_{s}M} \left(
p -\frac{1}{2p} \right) \sim \frac{2 \pi p}{3 g_{s}M} ~.
\end{equation}
Same as Eq.(\ref{dne}) in the slow-roll case, we have
\begin{equation}
-dN_{e} \simeq d \ln k \simeq d\ln l \simeq H dt ~.
\end{equation}
Suppose the $p$th step is at $l_{p}$ (or $\phi_{p}$),
because $\phi \simeq -\sqrt{\lambda_B}/t$, we get
\begin{equation}
\frac{\phi_{p}}{\phi_{p-1}}
\simeq \frac{t_{p-1}}{t_p}
\simeq \frac{{N_0 -\ln l_{p-1}}}{{N_0 -\ln l_{p}}} ~,
\end{equation}
where we have used the fact that $H$ is approximately a constant, and
$N_0$ is the number of e-folds (at the largest CMBR scale) to the
end of DBI inflation, for example $N_0 \approx 38$ \cite{Bean:2007eh}.
On the other hand, duality cascade relates the position of the adjacent steps
\begin{equation}
\frac{\phi_{p}}{\phi_{p-1}} = \exp\left(\frac{2\pi}{3g_s M} \right) ~.
\end{equation}
Suppose the feature at $l_{p_{i}+1}\sim 20$ is due to such a step, and suppose the feature at $l_{p_{i}}\sim 2$ is also due to a step (not just cosmic variance). That is, $p_{i}$ labels the initial or the first observable step.
Then, taking $N_0=38$, we see that the next 2 steps should be at
\be
l_{p_{i}+2} \sim 170 \qquad l_{p_{i}+3} \sim 1300 ~.
\ee
We also get a constraints on the microscopic parameters
\begin{equation}
g_s M \approx 33 ~.
\end{equation}

Now we estimate the value of $p_i$ by looking at the dip in the CMB
power spectrum at $l\sim 20$. The power spectrum is
\begin{equation}
P_\calR(k) = \frac{H^4}{4\pi^2 \dot\phi^2}.
\end{equation}
 In IR DBI inflation, the Hubble scale is dominated by the constant
$\sqrt{V_0}$, so it does not change across a step in the warp
factor. However, since $\dot \phi$ closely tracks the speed limit
$\dot \phi^2 \sim h^4(\phi)$, it is most sensitive to the step in the
warp factor. $\dot\phi^2$ increasing by a factor of $(1+2b)$ across a
step decreases $P_\calR$ by a factor of $(1+2b)$. The CMB data shows
that around $l\sim 20$, there is a dip in the power spectrum by about
$20\%$ in $k$-space, 
which gives $2b \sim 0.2$. Using (\ref{step_b}), we have
\begin{equation}\label{M_eq}
\left( \frac{3g_s M}{8\pi} \right) \frac{1}{(p_i+1)^3} \sim 0.2 ~.
\end{equation}
Given that $g_s M \approx 33$ to fit the spacing of the steps, we immediately get
\begin{equation}
p_i = 2 ~.
\end{equation}
With input of 4 quantities : the position $l \sim 20$, the fractional
height (size) $\Delta T/T \simeq 0.2$ and the width $\Delta l_{p} \sim
5$ of the $2$nd step as well as the position (at $l \sim 2$) of the
first step, we can use $\Delta l \propto l$ and $\Delta T/T \propto
p^{-3}$ to get a complete set of predictions:
\begin{center}
\begin{tabular}{cccc}
\hline
$p$ &   $l$ & $\Delta T/T$  & $\Delta l_{p}$ \\
\hline \hline
$2$ & $\sim 2$ & $\sim 0.7$ & $\sim 1$  \\
\hline
$3$ & $\sim 20$ & $0.2$ &$\sim 5$  \\
\hline
$4$ & $\sim 170$ & $\sim 0.08$ & $\sim 40$ \\
\hline
$5$ & $\sim 1300$ & $\sim 0.04$ & $\sim 260$ \\
\hline
\end{tabular}
\end{center}
Note that $\Delta T/T \sim 0.7$ and $\Delta T/T =0.2$ for the first
two steps are probably too big to be treated as a perturbation. We
shall take this to mean that the size of the first step can be very
big.

We should point out that the warp factor Eq.(\ref{step_b}) is a good
approximation only when $p \gg 1$. 
This condition is also necessary to ignore the running of the
dilaton.
So $p=3$ is already too
small. Moreover, as we
will see in Sec.~\ref{Sec:numeric}, the glitch in the power spectrum
around $l\sim 20$ may be
too large to be explained by a step feature in the warp factor due to
the duality cascades.

\subsection{Qualitative Analyses around a Single Step}
\label{Sec:Qualitative}
We consider a sharp step in the warped geometry $T(r)$. We parametrize
the size of the step as $2b \equiv \Delta T/T$ and the width as $\Delta
\phi \equiv 2d$. We use the $\tanh$ function interpolation between the two sides of the step, i.e.
\begin{equation}\label{step1}
T(r) \equiv T_3 \frac{r^4}{R^4} \left[1 + b\tanh \left( \frac{r-r_s}{d} \right) \right] ~.
\end{equation}

\subsubsection{The Evolution of the Sound Speed}
\label{Sec:csEvol}
We use the case of positive $b$ to illustrate the evolution of the
sound speed, the formulae are the same for negative $b$.
The direction of the step is such that, for positive $b$,
$T$ steps up as $r$
increases, i.e.~the warp factor $h \propto T^{1/4}$ (or the speed-limit
$h^2$) increases as $r$ increases.
 
The first stage is when branes move across the step, where
the speed-limit increases
suddenly.
Since the time that the branes spend to across the step is very short
comparing to the Hubble time (see Appendix \ref{AppWidth}), during
which we can ignore the Hubble expansion and the acceleration from the
potential. So across the step,
the behavior of the branes is
approximated by the Lagrangian $L=-T(r)\sqrt{1-\dot \phi^2/T(r)}$.
The corresponding energy $T \gamma = {\rm const}$, where $\gamma =
1/c_s = (1-\dot \phi^2/T) ^{-1/2}$.
Hence a sharp change in $T$ results in a sharp change in $c_s$, and
both changes are small,
\be
\frac{\Delta c_s}{c_s} = \frac{\Delta T}{T} = 2b ~.
\label{DcsDT1}
\ee
A small change in $c_s$ also means that the velocity of the branes
will closely track the change of the speed-limit in step.
 
The second stage is at the end of the step, where
the previous change in $c_s$ results in a deviation
from the attractor solution.
To see this, we note that in IR DBI inflation, the attractor solutions
have $\gamma \approx \beta
N_{e\rm DBI}/3$ and $N_{e\rm DBI} \approx HR/h$, so $c_s\approx
3h/\beta HR$. A fractional change in the warp factor $h=r/R$
means $\Delta R/R = -b/2$,
which implies that the attractor solutions has a fractional
change
\be
\frac{\Delta c_s}{c_s} = b ~.
\ee
So the increase in $c_s$
due to the first stage (\ref{DcsDT1}) over-shoots the new
attractor solution, the branes will have to quickly relax to the new
attractor solution. The Hubble expansion plays an important role here,
so the energy $T\gamma$ is no longer conserved.

\begin{figure}[th]
\begin{center}
\includegraphics[width=10cm]{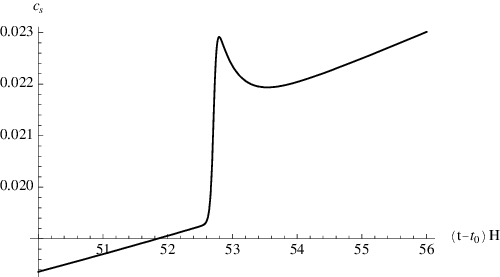}
\end{center}
\caption{\small Evolution of $c_s$. Parameters are $b=0.1$, step width $\Delta N_e=0.05$, 
$\beta=3$, $g_s
m_s^{-4}=10^{39}$, $N_B=10^9$, $n_B=10^4$, $n_Ah_A^4=16$.}
\label{Fig:cs}
\end{figure}

In summary, the behavior of $c_s$ can be approximated by two
step-functions side by side. In the first one, it jumps up from the
original $c_s$ to $c_s + 2b c_s$, during which the branes quickly
fall down
the step in warp geometry. The time scale is determined by $d$ and is
typically much smaller than $1/H$. The second one comes immediately
afterwards, it jumps down from $c_s + 2b c_s$ to $c_s + b c_s$, during
which the branes quickly approach the new attractor solution. The time
scale is of order $\CO(1/H)$.
The width of the second step function is much larger than the first
for narrower step (i.e.~smaller $d$).
Note that in both stages, the velocity of branes increases. This
evolution is illustrated in Fig.~\ref{Fig:cs}.

Because the width of step is very small, such a small change of $c_s$
happens in a short period. This gives rise to large $s\equiv \dot
c_s/(c_s H)$ and $\dot s/H$.
This turns out to be the primary sources to various observable
signatures in power spectrum and non-Gaussianities.

\subsubsection{The Power Spectrum}
\label{irdbi_pk}
The power spectrum is determined by the mode equation
Eq.(\ref{quadeom}).
We now estimate various parameters in Eq.~(\ref{z''/z}) 
in terms of the size, $b$, and width, $d$, of
the step feature in the warp factor. 
It is convenient to write $d$ in terms of 
the e-fold that the
brane spends moving across it,
\begin{eqnarray}
\Delta N_e \equiv H \Delta t \approx H \frac{d}{\dot \phi}
= \frac{d}{\sqrt{2c_s\epsilon}} ~.
\end{eqnarray}
Note this is not the e-folds that the feature on CMB spans (denoted
as $\Delta l$ previously) which
typically includes the oscillations and spreads much wider.

Using the continuity equation $\dot\rho = -3H(\rho + p)$, together
with the energy density and pressure Eq.(\ref{dbi_rho_p}), we can
solve for the evolution of $s$,
\begin{eqnarray}
s = \frac{\dot c_s}{H c_s} = 3 (1-c_s^2) + \frac{c_s \dot V}{T H}
+ \frac{\dot T}{TH} (1-c_s)
~.
\label{sExpression}
\end{eqnarray}
Here we can make more explicit
the conditions under which the sharp features in warp
factor dominate.
On the right hand side of (\ref{sExpression}), to make the 3$^{\mathrm{rd}}$ term
larger than the first, i.e.~to ignore the spatial expansion,
we require $3<\Delta T/ (T H \Delta t) \approx
2b/\Delta N_e$; to ignore the 2$^{\mathrm{nd}}$ term, i.e.~the sharp features in
potential, we require $c_s V' < T'$. Since the duality cascade leaves
sharp steps on $T(\phi)$ but not $V(\phi$), this condition is easily
satisfied.
Under these conditions, the last terms in the above equation
dominates, so we can approximate
\begin{eqnarray}
s \approx  \frac{\dot T}{T H} \sim
\frac{b}{d} \sqrt{2\epsilon c_s} \sim \calO \left(\frac{b}{\Delta
N_e}\right) ~.
\label{s_approx}
\end{eqnarray}
This is consistent with (\ref{DcsDT1}).

Denote $X\equiv \dot\phi^2/2$, so $\epsilon = X/(\Mpl^2 H^2 c_s)$
for DBI inflation. We get
\begin{eqnarray}
\tileta &=& \frac{\dot \epsilon}{H\epsilon} = \frac{\dot X}{H X} +
2\epsilon - s \ .
\label{relation1}
\end{eqnarray}
On the other hand, from the definition
$c_s = \sqrt{1-2X/T(\phi)}$,
we get
\bea
\frac{\dot X}{X H} = \frac{\dot T}{T H} - \frac{2c_s^2 s}{H(1-c_s^2)}
~.
\label{relation2}
\eea
Plug (\ref{relation2}) into (\ref{relation1}), and use
(\ref{sExpression}) to get rid of $\dot T/TH$, we have
\bea
\tileta &=& \frac{c_s}{1+c_s} s - 3(1+c_s) - \frac{\dot V c_s}{H T (1-c_s)}
+ 2\epsilon
\nonumber \\
&\approx& c_s s - (3 + \frac{\dot V c_s}{H T}) ~.
\label{etaExpression}
\eea
Except for the first term, the r.h.s.~of this expression is
affected little, $\CO(b)$, by the feature.
We use (\ref{etaExpression}) to further estimate $\dot
\tileta/H$. Keeping the time derivatives on $c_s$, $\dot \phi$ and
$T$, while ignoring those on $V'$ and $H$, we get
\bea
\frac{\dot \tileta}{H} &\approx& \frac{3}{2} s + c_s s^2 + c_s
\frac{\dot s}{H} ~.
\label{deta_approx}
\eea
We have used the fact 
that the brane always tracks the speed-limit,
\begin{eqnarray}\label{ddphi}
\dot \phi\approx \sqrt{T(\phi)} ~, \quad \ddot \phi \approx
\frac{1}{2}\frac{T'(\phi)}{T(\phi)}{\dot\phi}^2 ~,
\end{eqnarray}
since $c_s \ll 1$
even at the sharp feature as we know from Sec.~\ref{Sec:csEvol}.
Eq.~(\ref{s_approx}) is also used.

We further estimate $\dot s/H$,
\begin{eqnarray}
\frac{\dot s}{H} &=& \left(\frac{T'(\phi)}{T(\phi)}\right)'\frac{{\dot \phi}^2}{H^2} + \frac{T'(\phi)}{T(\phi)}
\left(\frac{1}{H}\frac{d}{dt}\frac{\dot \phi}{H}\right)  \nonumber \\
&\approx& \left(\frac{T''}{T} - \frac{T'^2}{T^2} \right) \frac{{\dot
\phi}^2}{H^2} + \frac{T'}{T}\frac{\ddot \phi}{H^2} ~. \label{ds/H}
\end{eqnarray}
Using Eq.(\ref{ddphi}) to eliminate the $\ddot\phi$ in Eq.(\ref{ds/H}), we get
\begin{eqnarray}
\frac{\dot s}{H} &\approx& 2c_s\epsilon \left(\frac{T''}{T} - \frac{1}{2}\frac{T'^2}{T^2} \right)
\approx 2c_s\epsilon \left(\frac{T''}{T}\right)
\sim b~ \frac{c_s\epsilon}{d^2} \sim \calO\left( \frac{b}{{\Delta N_e}^2}\right) ~,  \label{ds_approx}
\end{eqnarray}
where we have dropped the term $T'^2/T^2$, since $T''/T \gg T'^2/T^2$
as long as $b \ll 1$.

We now can compare the amplitudes of various terms. 
Because both $b$ and $\Delta N_e$ are small, from
(\ref{s_approx}) and (\ref{ds_approx}), we have
\bea
\frac{\dot s}{H} \gg s,s^2 ~.
\eea
From (\ref{etaExpression}) and (\ref{deta_approx}),
\bea
s \gg \tileta ~, ~~~\frac{\dot s}{H} \gg \frac{\dot \tileta}{H} ~.
\eea
So the only important contribution to $z''/z$ is from $\dot s/H$, and we have
\begin{eqnarray}
\frac{z''}{z} &\approx& 2a^2 H^2 \left(1 - \frac{\dot s}{2H} \right) \approx 2a^2 H^2
\left(1 - \frac{T''}{T} c_s\epsilon \right)~.
\end{eqnarray}
 
If we take $T$ to be the form Eq.(\ref{step1}), we can evaluate
\begin{eqnarray}
\frac{z''}{z} &\approx& 2a^2H^2 \left[ 1 - b \frac{c_s\epsilon}{d^2}\, \mathrm{sech}^2 \left(\frac{\phi-\phi_s}{d}\right)  \tanh \left(\frac{\phi-\phi_s}{d}\right) \right] \nonumber \\
&\sim& 2a^2H^2 \left[ 1 - \frac{b}{\Delta N_e^2}\, \mathrm{sech}^2 \left(\frac{\phi-\phi_s}{d}\right) 
\tanh \left(\frac{\phi-\phi_s}{d}\right) \right] ~.
\end{eqnarray}
The feature in $z''/z$ is dictated by the $\mathrm{sech}^2 \tanh$ term, and
will affect the evolution of certain perturbation modes $v_k$. As
moving across the step only generates $\sim N^{DBI}_e/M$
e-folds (See Appendix
\ref{AppWidth}), typically $b/\Delta N_e^2 \gg 1$. The mode $v_k$ will
first see a dip in $z''/z$ followed by a bump. Let us assume
that the feature in $z''/z$ shows up around conformal time $\tau_s$,
then modes with $c_s^2 k^2 \gg (z''/z) |_{\tau_s}$ or $c_s^2k^2 \ll
(z''/z) |_{\tau_s}$ will not be affected, because they are either
oscillating well inside the sound horizon or have already crossed the
sound horizon and got frozen. The major effect will be on the modes
with $c_s^2k^2 \sim (z''/z)|_{\tau_s}$. The range of $k$ affected by
the sharp feature is determined by $b/\Delta N_e^2$. If $P_\calR$
starts seeing the feature at $k_0$, the feature will disappear at
\bea
\Delta k \sim k_0 \sqrt{b}/\Delta N_e ~.
\label{kextDBI}
\eea

\subsubsection{Non-Gaussianities}
\label{Sec:nonGsharpDBI}
Although Ref.~\cite{Chen:2006nt,Maldacena:2002vr,Seery:2005wm} are
only interested in non-Gaussianities without sharp features, the cubic
expansion of the perturbations is exact and does not rely on the
assumption that various inflationary parameters $\epsilon$, $\tileta$
and $s$ are small.
For DBI inflation, among all the terms in Ref.~\cite{Chen:2006nt}, 
the leading term in the cubic action that is responsible for sharp
features in non-Gaussianity is
\begin{eqnarray}
\frac{a^3 \epsilon}{2 c_s^2} \frac{d}{dt} \left( \frac{\tileta}{c_s^2}
\right) \zeta^2 \dot \zeta ~.
\label{cubicterm}
\end{eqnarray}
In the absence of sharp features, a term such as
$(a^3\epsilon/c_s^4) \zeta \dot \zeta^2$ contributes to the
non-Gaussianity estimator $f_{NL} \sim 1/c_s^2$. Therefore as a rough
estimate the term (\ref{cubicterm}) contributes
\begin{eqnarray}
f_{NL}^{\rm feature}
\sim \frac{d}{dt} \left( \frac{\tileta}{c_s^2} \right)
\Delta t  \sim \Delta \left( \frac{\tileta}{c_s^2} \right)
~.
\end{eqnarray}
The cases with large $s$ are most interesting for
non-Gaussianities. 
Using (\ref{etaExpression}),
\bea
\tilde \eta \approx c_s s ~,
\label{etacs}
\eea
so
\bea
f_{NL}^{\rm feature} \sim \frac{\Delta s}{c_s} 
\sim \frac{1}{c_s} \frac{b}{\Delta N_e} ~.
\label{fNLIRDBIfeature}
\eea
There is another term
\bea
-2\frac{a\epsilon}{c_s^2} s \zeta (\partial \zeta)^2
\eea
in the cubic action of Ref.~\cite{Chen:2006nt} that contributes 
\bea
f_{NL}^{\rm feature} \supset \frac{s}{c_s^2} \Delta t H
\sim \frac{\Delta c_s}{c_s^3} \sim \frac{b}{c_s^2} ~,
\eea
which is also possibly observable. The term (\ref{fNLIRDBIfeature})
dominates for the most interesting cases.

The net observable effect will be a nearly-scale-invariant large
non-Gaussianity of order $\CO(1/c_s^2)$ plus the scale-dependent
(oscillatory) modulation of order $\CO(\Delta s/c_s)$.
Similar to the slow-roll case, we can write $f_{NL}^{\rm feature}
\sim (\sqrt{b}/c_s) (\sqrt{b}/\Delta N_e)$. 
As we have shown, the first
factor
determines the amplitude of the glitch in power spectrum through $b$, 
while the
second factor determines the extension of its oscillations through
(\ref{kextDBI}). So there is a simple relation between 
the effects of sharp feature on the power spectrum
and non-Gaussianity. For a same single visible glitch in power spectrum,
the oscillatory 
amplitude of the bispectrum in DBI inflation is
larger than that in slow-roll inflation
mostly due to the factor of $1/c_s$ which can make $f_{NL}^{\rm
  feature}$ 
larger than $\CO(10)$; while for glitches that have
long oscillatory extension (which can also be made invisible in
power spectrum with tiny $b$), 
the associated bispectrum can be very large with tiny $\Delta N_e$.
The estimates of $b$ and $\Delta N_e$ in Appendix \ref{Appwarp} and
\ref{AppWidth} suggest that the well-separated steps in the duality cascade
tend to fall into the latter category. 
We will later see independent evidence from fitting the power spectrum
glitch to WMAP CMBR data.

\subsection{An Analytical Approximation for a Single Sharp Step}
\label{Sec:step_ana}
In this subsection, we give some analytical treatment of the effect of
sharp feature on power spectrum. The full analyses rely on numerical
calculations. Nonetheless we can see how several properties emerge and
we summarize them in the end of this subsection.

In Section \ref{Sec:csEvol}, we see that the sound speed undergoes two
adjacent jumps. Here we simplify it by just considering the overall
net effect, namely we approximate it by one step-function from $c_s$
to $c_s = b c_s$. We approximate the transition to be infinitely sharp
so that an analytical analysis is possible.
As we showed, the width of the first step function can be made 
very small as we
reduce $d$, while that of the second is much wider and of order
$\CO(1/H)$ model-independently. So this approximation is only 
qualitative.

We denote $\tau_0$ to be the moment inflaton passes the feature, 
and look at the
solutions of the equation of motion (\ref{quadeom}) at both sides of
$\tau_0$.
At $\tau = \tau_0^-$,
\begin{eqnarray}
v_k = v_1 |_{\tau_0^-} ~.
\label{vk-}
\end{eqnarray}
At $\tau = \tau_0^+$,
\bea
v_k = C_1 v_1|_{\tau_0^+} + C_2 v_2 |_{\tau_0^+} ~.
\label{vk+}
\eea
Here we have denoted the two linearly independent solutions as
\bea
v_1 &\equiv& \frac{i a H}{\sqrt{2c_s^3 k^3}} (1+i k c_s \tau)
e^{-i k c_s\tau}
~, \\
v_2 &\equiv& \frac{i a H}{\sqrt{2c_s^3 k^3}} (1-i k c_s \tau)
e^{i k c_s\tau} ~.
\eea
For $\tau<\tau_0$ we only have $v_1$ because the vacuum is dominated
by the Bunch-Davis vacuum.
 
These two solutions are matched according to the boundary conditions
at $\tau_0$. These are contributed dominantly by the sharp change in
$z''/z$ in the last term of (\ref{quadeom}).
From the previous subsection, we know that 
the change in $z\equiv a\sqrt{2\epsilon}/c_s$
is dominated by the change in $c_s$. As
mentioned we approximate
\bea
c_s = c_s|_{\tau_0^-} + b c_s|_{\tau_0^-} \theta(\tau-\tau_0) ~,
\eea
where $\theta$ is the step-function.
So
\bea
z=\left\{
\begin{array}{lr}
z_-(\tau) ~, & \tau<\tau_0 ~,\\
z_0 -b z_0 \theta(\tau-\tau_0) ~, & \tau=\tau_0 ~, \\
z_+(\tau) ~, & \tau>\tau_0 ~,
\end{array}
\right.
\eea
where $z_0=z_-(\tau_0)$. Around $\tau_0$, we then have
\bea
v_k'' &\approx& \frac{z''}{z} v_k ~, \\
&\approx& \left[ -b \frac{d}{d\tau} \delta(\tau-\tau_0)
+ \frac{\Delta z'}{z_0} \delta(\tau-\tau_0) + \cdots \right] v_k ~,
\label{vkboundary}
\eea
where the dots denotes terms that do not contribute upon integration
over the infinitely small region around $\tau_0$, and $\Delta z'
\equiv z'|^{\tau_0^+}_{\tau_0^-}$.
Because $z' = aH z (1+\eta/2-s)$, we have
$z'|_-^+ = aH z|_-^+$.
Integrating (\ref{vkboundary}) once and twice, we get the two boundary
conditions for $v_k$,
\bea
v_k|_{\tau_0^+} - v_k|_{\tau_0^-} = -b v_k|_{\tau_0^-} ~,
\label{bcond1}
\eea
and
\bea
v'_k|_{\tau_0^+} - v'_k|_{\tau_0^-} = -b a H v_k|_{\tau_0^-} ~.
\label{bcond2}
\eea
 
Matching (\ref{vk-}) and (\ref{vk+}) we get
\begin{eqnarray}
C_1 &=& \left[1 - \left(\frac{1}{2} + \frac{i}{2x_0}\right)b - \left(\frac{1}{8} - \frac{5i}{4x_0}\right) b^2 \right] e^{ibx_0} \\
C_2 &=& \left[\frac{i b}{2 x_0} + \left(\frac{1}{2} - \frac{5i}{4x_0}\right) b^2 \right]e^{-i(2+b)x_0}
\end{eqnarray}
where we denote $x_0 = \tau_0 kc_s|_{\tau_0^-}$.
The power spectrum is
\begin{equation}
P_\calR = \frac{k^3}{2\pi^2} \frac{|v_k|^2}{z^2} \Bigg|_{\tau\to 0}
= P_{\calR 0} \frac{|C_1+C_2|^2}{1+b}
\label{PowerSharp}
\end{equation}
where $P_{\calR 0}={H^2}/(8\pi^2 \epsilon c_s|_{\tau_0^-})$. For small $k$,
$x_0\to 0$, so $P_\calR \to P_{\calR 0}$ as expected. For large $k$, $P_\calR$ is
oscillating. The fact that such oscillation extends to infinitely
large $k$ is the artifact of approximating the change in $c_s$ to be
infinitely sharp. Such an approximation makes the potential barrier $z''/z$ infinitely high,
and introduces the $v_2$ component to all $k$ modes.
Realistically, the relaxation takes finite time $\Delta N_e$, and the potential barrier in $z''/z$ is finite.
For modes with high frequency $kc_s \gg \sqrt{z''/z}$, it will not see the barrier in $z''/z$ and
the $v_2$ component is absent. Therefore, for the mode
\begin{equation}
k \gg \sqrt{\frac{z''}{z}} \sim k_0 \frac{\sqrt{b}}{\Delta N_e}
\end{equation}
the oscillation is damped away.
Here $k_0 \equiv aH/c_s$ is the mode number
crossing the horizon at $\tau_0$.
 
\begin{figure}[th]
\begin{center}
\includegraphics[width=12cm]{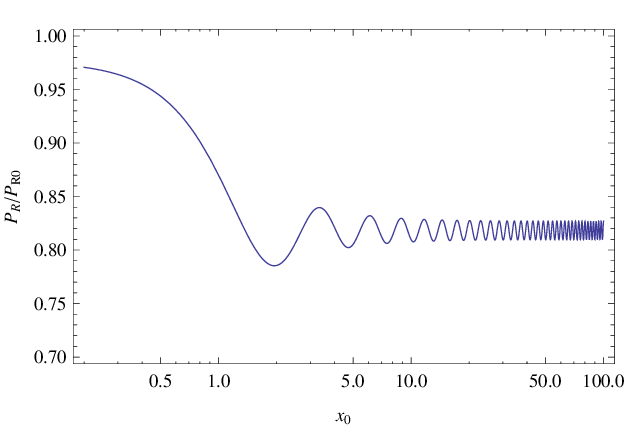}
\end{center}
\caption{\small The approximate
power spectrum, $P_{\calR}$, as a function of $x_{0}=kc_s\tau_0$, where $\tau_0$ is the time of the feature, due to a $\theta$ function step with $b=0.1$. }
\label{Fig:Pksharp}
\end{figure}
 
This power spectrum is illustrated in Fig.~\ref{Fig:Pksharp}. The
fact that the dip appears first for positive $b$ can be seen
analytically by taking the $bx_0\ll 1$ limit,
\begin{equation}
P_\calR \to P_{\calR 0} \left( 1-2b+\frac{b}{x_0} \sin 2x_0 \right) ~.
\label{pk_sharp}
\end{equation}

Here we summarize several general properties from this
analytical approximation. Firstly, the oscillation behavior in power
spectrum can
be understood as being due to the change of vacuum caused by the sharp
feature. Such a change introduces a second (negative energy)
component of $v_k$ in addition to the original Bunch-Davis (positive
energy) component. The superposition results in a oscillation in
k-space. Secondly, the matching of the two set of solutions are done
at the location of the sharp feature $x_0 = kc_s\tau_0 = k/k_0$, where
$k_0\equiv (-\tau_0 c_s)^{-1}$
is the wave number crossing the horizon at
the moment of the feature $\tau_0$.
The phase $2x_0$ gives the oscillation wavelength
$\sim k_0/2$ in power spectrum model-independently.
This implies that, in $\log k$ coordinate,
the oscillation behavior looks similar for sharp features at
different $k_0$.
Lastly, for a feature
with finite width, the high frequency modes in the vacuum change
adiabatically and remain in the Bunch-Davis vacuum. So the effect of
the feature decay away for those modes. These conclusions apply for
both the DBI and slow-roll case.
 
\subsection{Numerical Analyses and Data Fitting}
\label{Sec:numeric}
 
\begin{figure}[t]
\begin{center}
\includegraphics[scale=0.7]{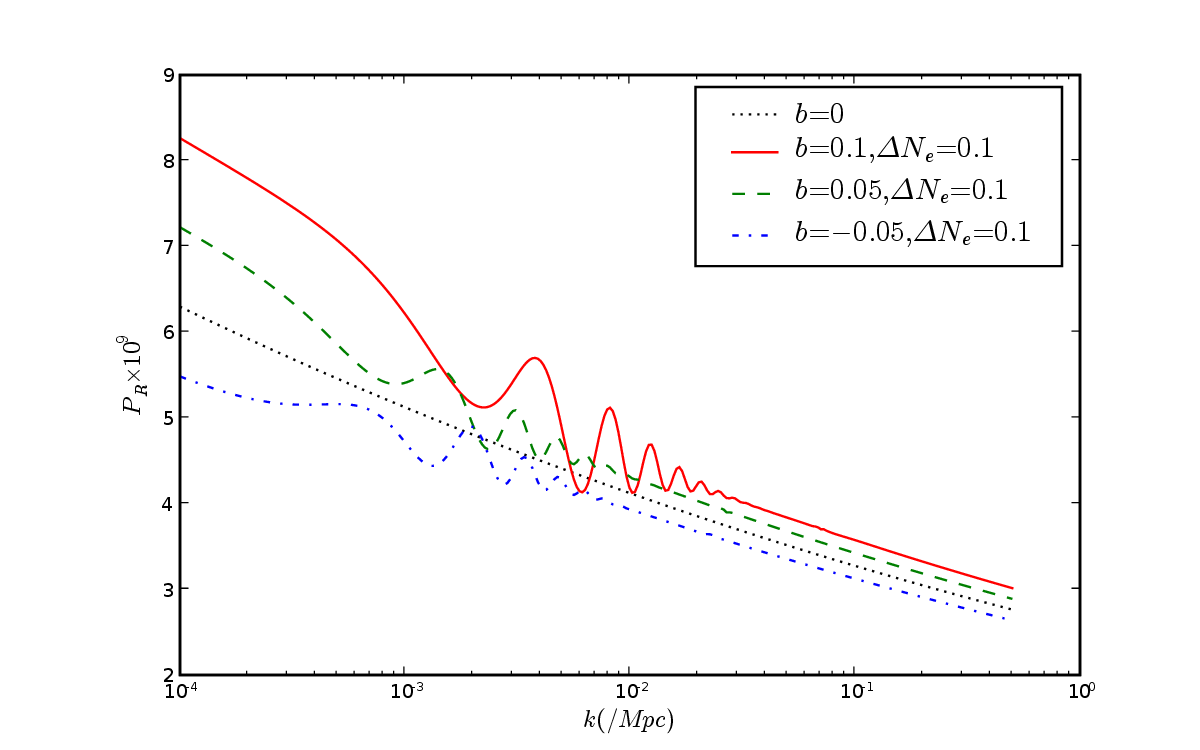}
\caption{\small 
In the IR DBI scenario, we show the power spectrum, $P_{\calR}$, when there is a
sharp step in the warp factor. For the same step width $\Delta N_e$,
we show three cases with different step size $b$. The amplitude of the
first dip and bump increases as we increase the step height $b$. A
bump appears first for $b<0$ while a dip appears first for $b>0$. }
\label{fig:step_b}
\end{center}
\end{figure}

\begin{figure}[t]
\begin{center}
\includegraphics[scale=0.7]{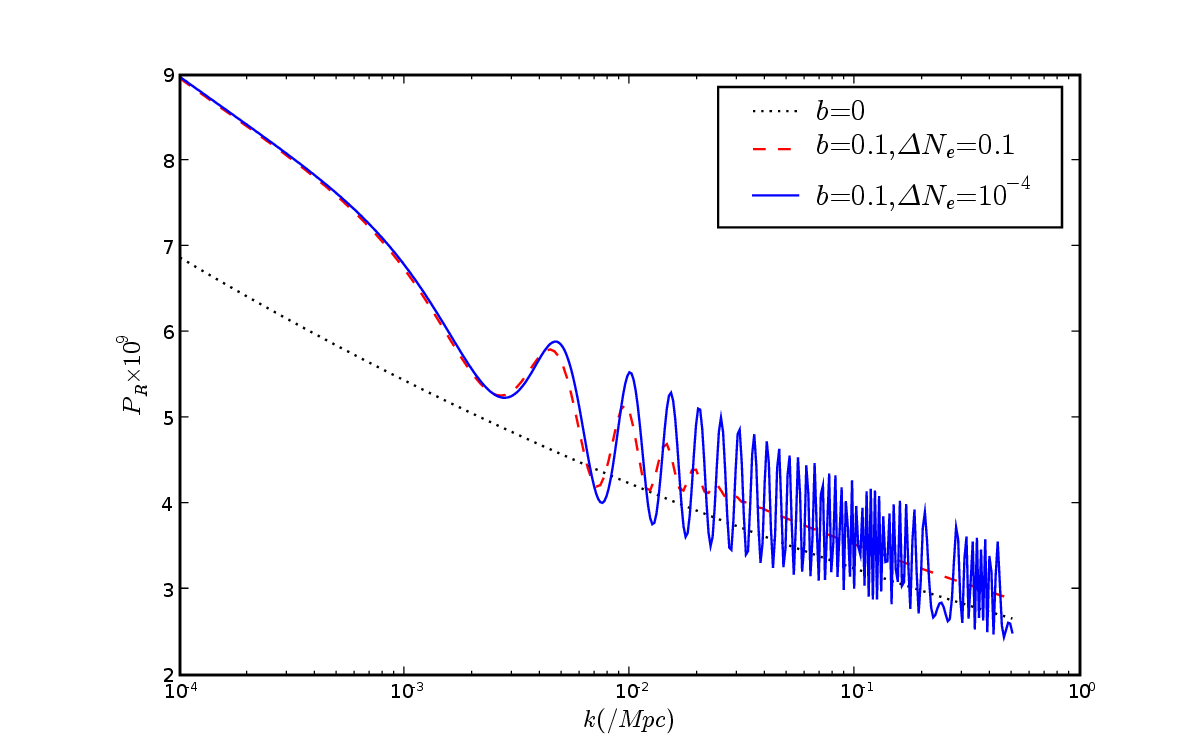}
\caption{\small Same as Fig.~\protect\ref{fig:step_b}, with the step size $b$ fixed,
we show two cases with different step width $\Delta N_e$. Since in IR DBI model, the typical 
step width corresponds to $\Delta N_e \ll 1$, the dip and bump in the power spectrum is not 
sensitive to $d$. The width of the step affects only the range of
oscillation in $P_\calR$.} 
\label{fig:step_d}
\end{center}
\end{figure}

We have analyzed qualitatively the behavior of the power spectrum around the step. However, due to 
the non-trivial behavior of $z''/z$, the mode equation is hard to solve analytically in general. To better 
understand the effect of the sharp feature,
we numerically solve the equation of motion Eq.(\ref{dbi_eom}), evolve
the mode equation Eq.(\ref{quadeom}) starting from when the mode is
deep inside the sound horizon ($c_s^2k^2 \gg z''/z$) until the mode is
frozen when stretched out of the sound horizon ($c_s^2k^2 \ll z''/z$),
and evaluate the power spectrum numerically using
Eq.(\ref{powerspec}).
The numerical calculation also has the advantage that it is not
necessary to assume certain simplification conditions in
Sec.~\ref{Sec:Qualitative}, such as $b/\Delta N_e \gg 1$.
Our major results are shown in Fig.~\ref{fig:step_b} and Fig.~\ref{fig:step_d}.

Fig.~\ref{fig:step_b} shows the step feature in the power spectrum
$P_\calR$ for steps with different height $b$, but same width $\Delta
N_e$. We see clearly that in the power spectrum, a dip appears first
with positive $b$ and a bump appears first with negative $b$, which
agrees with our analytic result Eq.(\ref{PowerSharp}). However such a
sequence gets less clear after projecting from $k$-space to
$l$-space.

In Fig.~\ref{fig:step_d} we explore the step feature with different
step width $\Delta N_e$. We see that the oscillation amplitude in
$P_\calR$ is quite insensitive to the step width $d$. 
By the same argument in our slow-roll analysis, the steps are sharp so
that $\Delta N_e \ll 1$. The $c_s$ and $\dot c_s$ have reached their
maximum values and so is the amplitude of the oscillation which is
controlled by $b$ as in (\ref{PowerSharp}).
 
\begin{figure}[th]
\begin{center}
\includegraphics[scale=0.6]{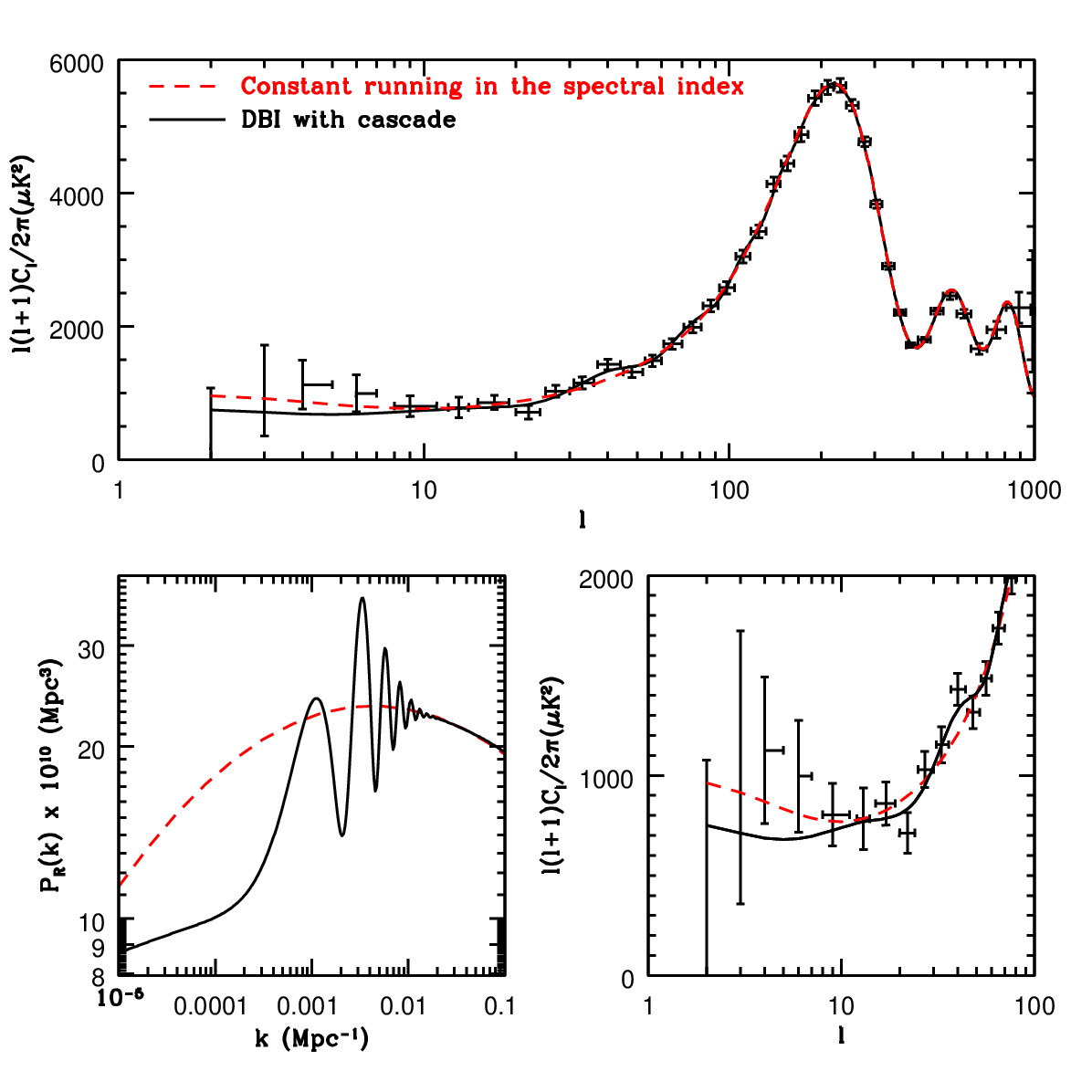}
\caption{\small The CMB temperature power spectrum (top) and initial power
spectrum (bottom left) in the presence  of a step in the warp factor
(full black line)  and a best-fit scenario with constant running in
the spectral index (red dashed line). The presence of large scale
features in the cascade spectrum (bottom right) lead to an improved
fit with the WMAP 3-year data of $\Delta\chi^2 =-2.76$ over the model
with constant running in the spectral index.}
\label{Cls}
\end{center}
\end{figure}

In Fig.~\ref{Cls} we perform a $\chi^2$ fit to the WMAP data. With
one single step in the warp factor, we try to fit the $l \sim 20$ dip
in WMAP temperature anisotropy. Here, instead of performing a full
MCMC search for the best fit model, we only show an example to
illustrate how the local step feature improves the quality of data
fit. The example we show in Fig.~\ref{Cls} has the IR DBI model
parameters $n_B = 6761, N_B = 4.315\times 10^{9}, n_Ah_A^4 =
0.01035, m_s/g_s^{1/4}  = 2.567\times 10^{-9}M_p, \beta = 4.021$, and
the step parameters are $\Delta N_e = 0.105$ and $b=-0.35$. The step
model, with  likelihood ${\mathcal L}$, $-2\ln {\mathcal L} =5347.16$
when compared to WMAP 3-year temperature and polarization data
\cite{Hinshaw:2006ia,Spergel:2006hy,Page:2006hz,Jarosik:2006ib}, has a
better fit to the data, $\Delta\chi^{2}=-2.76$, over the best-fit
scenario with constant running of the spectral index, also shown. In
light of the four extra degrees of freedom\footnote{Three of them
  specify the location, height and width of the step, another one is
  from IR DBI (versus LCDM).}
determining the cascade
power spectrum in comparison to the fiducial running model, however,
this improvement in $\chi^2$ is not statistically significant.
 
We also find that a positive $b$ is not favored by data, so the first
dip in the $k$-space does not necessarily transform into a clear dip
in the $l$-space.
Numerically
we find that to fit the $l\sim 20$ dip, we need $|b|\sim 0.3$.
However, if we take $b=+0.3$, it gives too much power on large
scale, more than that is allowed by data. 

The reason that the glitch appears to be less sharper than that in
slow-roll case is the following. In DBI inflation, a step in warp
factor not only causes oscillation in power spectrum, but also changes
the asymptotic speed-limit of the inflaton. So in addition to glitches,
it also introduces a step in the power spectrum. Both amplitudes are
controlled by the relative height of the step $b$. It is more
difficult to generate a sharp glitch while keeping the latter effect
compatible with data. 
While in slow-roll case, asymptotic velocities of the inflaton is not
affected by a local step in potential.
In this sense, for DBI inflation the glitches in power spectrum can
be made sharper if one introduces a bump (or anything that does not change
the asymptotic behaviors of the warp factor) instead of a step in the
warping.

We should also point out
that $|b|\sim 0.3$ gives a step much larger than that given by Eq.(\ref{step_b})
\footnote{It is possible that duality cascade will
generate large steps if we take $p\sim 1$. However, the perturbative
analysis of the warp factor breaks down with $p\sim 1$, and we do not
have any analytic control in that regime.}, which typically goes like
$1/p^3$ with $p \gg 1$.
This means that if we take the steps generated
by the duality cascade, most probably we will not see any observable
effects in the power spectrum. However, as we estimated in
Eq.~(\ref{fNLIRDBIfeature}), the non-Gaussianities associated with the
step can be very large depending on parameters in 
Eqs.~(\ref{Dh/h}) and
(\ref{NeSD}). This can leave distinctive features detectable by experiments.

\subsection{Closely Spaced Steps}
\label{Sec:Res}

We have been mainly concentrated on features that are sharp enough
and well-separated (by at least one e-fold).
We also see that there are parameter space
where the feature is not sharp enough to show up in the CMB as
observable signatures. In this case, the step is no longer important
individually; but if there are parameter space such that, within one
e-fold, the branes can go over many such small steps, a different
kind of large
non-Gaussianities can be generated through the ``resonance mechanism''
\cite{Chen:2008wn}. These closely spaced small steps induce small but 
high frequency oscillations of the background evolution in
$\epsilon$, $\tilde \eta$ and $s$. This oscillation can resonant with the mode
functions $u_k$ when the mode is still within the horizon. As a result
of this resonance, the non-Gaussianity integration picks up a large
contribution while the two-point function is only slightly affected.
This non-Gaussianity has different signatures from the sharp feature
case \cite{Chen:2008wn}.

In slow-roll model, the condition to achieve this resonance mechanism
relies on the shape of the potential. Naively, since we already have a series
of steps, we can increase the tilt of the potential to make the branes
roll faster so they come across many steps in one
e-fold. Realistically, one needs to make sure that such a tilt will not
significantly reduce
the already-tuned total number of e-folds.

In IR DBI model, as we have discussed at the beginning of
Sec.~\ref{Sec:irstep}, there is actually a quite natural parameter
space that such a condition can be satisfied, $1>g_s>10^{-3}$. We
leave a more detailed study to a future publication.

\section{Remarks}
\label{Sec:summary}
The analysis of the step feature in inflation in comparison with WMAP
data has been carried out for large field models, specifically the
quadratic chaotic inflationary model where the inflaton field starts
at values much bigger than the Planck mass ($\phi > 15 \Mpl$) and then
decreases to zero. In brane inflation, the inflaton being the position
of the brane, is bounded by the size of the flux compactification
volume, and so typically $\phi \ll \Mpl$. This requires a generalized
analysis, which is carried out here. In the IR DBI model,
the behavior of the inflaton is further modified due to the DBI
kinetic term. We have studied the properties and the signatures of the
steps in warp factor and see that typically the effect on the power
spectrum is much smaller, but non-Gaussianities can be much larger.
So non-Gaussianities becomes a more important tool to probe the sharp
features in DBI inflation.
Another likely scenario is when the steps are too close to show up as
sharp features
in the power spectrum. Their presence is expected to show up in the
bi- and tri-spectra.
It is important to study this case further.

We end with some remarks on two broader aspects on brane inflation.
In this paper we considered brane inflation as an effective
one-field model driven by a potential in the radial direction.
The location of the brane in each of the compact directions (one
radial and five angular) is represented by a scalar field -- brane
inflation is therefore multi-field in nature.  Multi-field
models can be decomposed into adiabatic and isocurvature fields,
respectively describing motion along, and perpendicular to, an
evolutionary trajectory in field space, where sharp features, such as
sudden turns, in the trajectory can give rise to the interconversion
of adiabatic and isocurvature modes
\cite{Lyth:2006nx,Vernizzi:2006ve,Huang:2007hh}. 
The introduction of isocurvature perturbations can significantly modify
the primordial power spectrum, and hence tight constraints on their
contributions can be placed from observed CMB and large scale
structure spectra \cite{Bean:2006qz,Keskitalo:2006qv}.
It will be 
interesting to see if these effects can be realized in brane inflation
models.

In brane inflation, there are a number of possible sources of non-Gaussianities:
(1) Even for a single $D3$ brane case, the inflaton is actually a six component field, namely, 
the radial mode plus the five angular modes. The angular modes are
massless in the simplest scenario, though in general they are expected
to pick up small masses.  This is a multi-field inflation, where
non-Gaussianity may arise; (2) the DBI action; (3) cosmic strings and
their cosmic evolution, and (4) step-like behavior. In this paper, we
focus on the step-like behavior due to the warp geometry, and its
possible detection. If more than one source of non-Gaussianity are
present, it will be important to disentangle them, for example via
bi-and tri-spectra. In this sense, brane inflation can be very rich.

\acknowledgments
We thank Hiranya Peiris for providing the MCMC chain.
We thank Richard Easther, Hassan Firouzjahi, Eiichiro Komatsu, Eugene Lim, Liam McAllister, Sarah Shandera, Gary Shiu and Eva Silverstein for valuable discussions.
XC, HT and JX would like to thank the Kavli Institute for Theoretical
Physics in China and the organizers of the ``String Theory and
Cosmology'' program for warm hospitality.
The work of RB is supported by the National Science Foundation under grants AST-0607018 and PHY-0555216. 
XC is supported by the US Department of Energy under cooperative
research agreement DEFG02-05ER41360.
The work of GH, SHT and JX is supported in part by the National Science Foundation under grant PHY-0355005.

\appendix
\section{The Warp factor}
\label{Appwarp}
Let us consider the structure in the warp metric $h(r)$ that rescales masses, where
$$ds^2=h^{2}(r)(-dt^{2} + a(t)^{2}d{\bf x}^{2}) + h^{-2}(r)(dr^2+r^2 ds_{5}^2)$$
For the KS throat, we have the approximate Klebanov-Tseytlin solution \cite{Klebanov:2000nc}
\be
h^{-4}(r) =  \frac{27 \pi {\ap}^{2} g_{s}}{4 r^{4}} \left( N + \frac{3g_{s}M^{2}}{2 \pi} [\ln (r/R) +1/4] \right),
\label{warpH0}
\ee
where, with the volume of $T^{1,1} \sim S^3 \times S^2$ given by $v \pi^{3}=16\pi^{3}/27$,
\be
R^4 = 4 \pi \alpha'^{2}g_{s}N/v = \frac{27 \pi \alpha'^{2}g_{s}N}{4}
\ee
and $M$ is the RR-flux wrapping $S^3$ while $K=N/M$ is the orthogonal NS-NS-flux.
The locations of duality transitions in the KS throat are given by
\be
r_{l}= R \exp\left(-\frac{2l\pi}{3g_s M}\right).\label{rlr0fa}
\ee
Note that the constant piece $N$ in (\ref{warpH0}) is determined by the boundary condition.
We see that the effective $D3$-brane charge is given by
\be
N_{\mathrm{eff}} = N + \frac{3g_{s}M^{2}}{2 \pi} \ln (r/R)
\ee
so that at $r=r_{l}$, $N_{\mathrm{eff}} = N - lM=K-l$. The term with $1/4$ factor is introduced to ensure that the warp factor $h^4(r)$ is monotonic for $R \ge r \ge r_{K}$.

The size of the step at $r=r_{p}$ in the warp factor $h(r)$ for large $p < K$ (where $r=r_{K+1}$ is at the edge of the throat) is estimated perturbatively to be (with string coupling $g_{s}$
valued at the step),
\be
\frac{\Delta h(r_{p})}{h(r_{p})} = \frac{ h_{>}(r_{p}) -  h_{<}(r_{p})}{h(r_{p})} \simeq \left(\frac{3g_s M}{2\pi}\right)\frac{1}{p^3}.
\label{Dh/h}
\ee
where $h_{>}(r_{p})$ is the warp factor at $r \ge r_{p}$ and $h_{<}(r_{p})$ that at $r \le r_{p}$.
The widths of the steps are estimated in Appendix \ref{AppWidth}. This
series of steps leads to a cascading feature in the warp factor. So,
in general, we expect this multi-step feature to be generic; as we
approach the infrared (decreasing $r$), the step size grows (as $p$
decreases). The spacings between steps are roughly equal as a function
of $\ln r $ for large $p$.

In another throat, it is entirely possible that the leading ${\Delta h(r_{p})}/{h(r_{p})}$
appears at a different order, say $p^{-2}$ instead of $p^{-3}$. It is also possible that ${\Delta h(r_{p})}/{h(r_{p})}$ can be negative, instead of positive, so $h(r)$ is ``saw-like'' instead of a cascade.
 
Here we write down the warp factor with steps given in \cite{Hailu:2006uj} for a flow including only $s$ number of steps starting from some location of duality transition at $r=r_{l_0}$; i.e, in the range $r_{l_0-1}<r<r_{l_0+s}$.
The warp factor in this case can be written as
\begin{eqnarray}
h^{-4}(r) & \simeq & h^{-4}(l_0,r)
  + \frac{27\pi \alpha'^2}{4 r^4}\sum_{l=l_0}^{l_0+s-1}\Bigl[g_s p_l M\left(\frac{3g_s M}{8\pi}\right)\nonumber\\&  &\times \frac{1}{2p_l^3}
\Bigl(1+\tanh [ \frac{r_{l}-r}{q r_{l}} ]\Bigr)\Bigr],\label{Hrdef2}
\end{eqnarray}
where $p_l\equiv K-(l-1) $ and $h^{-4}(l_0,r)$ is given by
\begin{eqnarray} {{h^{-4}}(l_0,r)}& \simeq & \frac{27 \pi \alpha'^2}{4 r^4}\Bigr[
g_s N+\frac{3 g_s^2 M^2}{2\pi}  [\ln
   (\frac{r}{r_0})+\frac{1}{4}]\nonumber\\&&-
   \Bigl(\frac{3 g_s^2 M^2}{2\pi}[\ln(\frac{r}{r_{l_0}})+\frac{1}{4}][\frac{3g_s M}{\pi}
   \ln(\frac{r_{l_0-1}}{r_0})+(2l_0-3)-\frac{3g_s M}{8 \pi}]\nonumber\\&&-\frac{9 g_s^3 M^3}{4\pi^2}  [\ln
   (\frac{r}{r_0})+{2}(\ln(\frac{r}{r_0}))^2+\frac{1}{4}]
   \Bigr)\frac{1}{K-(l_0-1)}\Bigr].\label{HdirDE}
\end{eqnarray}
Note that the size of the step in $h^{-4}(r)$ at $r=r_l$ is
\begin{eqnarray} {h^{-4}}(l,r_l)-h^{-4}(l+1,r_l)& \simeq & -
\frac{27\pi \alpha'^2 g_s N_{\mathrm{eff}}}{4 r_l^4}\left(\frac{3g_s M}{8\pi}\right) \frac{1}{p_l^3},\label{Hdiffstepl}
\end{eqnarray}
where $N_{\mathrm{eff}}=N-(l-1)M=p_l M$ is the effective $D3$-branes charge for the flow from the $(l-1)^{th}$ to the $l^{th}$ duality transition point.

\section{The $z''/z$ in Slow Roll Case}
\label{AppRelax}
In the presence of a feature like the step in the potential, the
motion of the inflaton is divided into three phases: attractor before
the step, acceleration within the step, and relaxation after the
step. During the attractor phase, the inflaton moves along with the
slow-roll attractor determined by the flat potential.  In the
acceleration phase, the inflaton gets accelerated by the local steep
potential due to the step, and deviate away from the attractor
solution. In the relaxation phase, the potential is flat again and the
inflaton motion relaxes back to the slow-roll attractor due to Hubble
friction.

Since Hubble friction is responsible for damping the deviation from
slow-roll attractor, it is natural to expect that the relaxation time
for $\dot\phi$ is of order $H^{-1}$ or ${\cal O}(1)$ e-fold, and the
same for $\epsilon$, $\tileta$. However, the sharp change in $z''/z$ 
takes place within a period much shorter than ${\cal O}(1)$ e-folds. To see
this, we first express $z''/z$ in terms of inflation parameters,
\begin{eqnarray}
\frac{z''}{z} &=& 2a^2 H^2 \left( 1 - \frac{1}{2}\epsilon - \epsilon^2 + \frac{3}{4}\tileta + \epsilon\tileta + \xi^2 \right) \nonumber \\
&\approx& 2a^2 H^2 \left( 1 + \frac{3}{4}\tileta + \xi^2 \right) ~,
\label{z''/z_sr_def}
\end{eqnarray}
where in the last line we have ignored the $\epsilon$ term since $\epsilon \ll 1$ in small field brane inflation models. The definitions of $\epsilon$ and $\tileta$ follow the main text and $\xi^2$ is defined the following way
\begin{eqnarray}
\xi^2 \equiv \frac{d^3\phi}{dt^3} \frac{1}{2H^2\dot\phi}  \ .
\end{eqnarray}
We can use the following equation of motion to eliminate the term
$d^3\phi/dt^3$ in $\xi^2$,
\begin{eqnarray}
\ddot{\phi} + 3H\dot\phi + \frac{dV}{d\phi} = 0 ~.
\end{eqnarray}
We then get
\begin{eqnarray} \label{xi2}
\xi^2 = -\frac{V''(\phi)}{2H^2} + 3\epsilon - \frac{3}{4}\tileta
\approx -\frac{V''(\phi)}{2H^2} - \frac{3}{4}\tileta ~.
\end{eqnarray}
Combining (\ref{z''/z_sr_def}) and (\ref{xi2}), we get
\begin{eqnarray} \label{z''/z_sr_simp}
\frac{z''}{z} \approx 2a^2H^2 \left(1 -\frac{V''(\phi)}{2H^2} \right)
~.
\end{eqnarray}
The result (\ref{z''/z_sr_simp}) is very useful in determining the
relaxation time of $z''/z$. We see that the sharp feature in $z''/z$
vanishes once $V''(\phi)$ goes back to normal, which only requires
that $\phi$ has moved across the step. Since the potential is very
steep at the step, it only takes $\Delta N_e \ll 1$ e-folds for $\phi$
to move across. So $z''/z$ relaxes much faster than $\epsilon$,
$\eta$ and $\xi^2$. When $z''/z$ goes back to the attractor behavior,
$\epsilon$, $\eta$ and $\xi^2$ still needs a few more e-folds to
relax. Both are important to observational quantities.

\section{Estimate the width of the steps}
\label{AppWidth}
\subsection{The width from Seiberg duality}
The running of the two coupling constants $g_1$ and $g_2$ behave as
\bea
\frac{1}{g_{1,2}^2} = \pm \frac{3M}{8\pi^2} \ln \frac{\Lambda}{\mu}
+ \frac{1}{g_0^2} ~.
\label{grunning}
\eea
The constant $g_0$ and the scale $\Lambda$ are chosen such that at
$\mu=\Lambda$, $g_1=g_2 \equiv g_0$. In terms of the string coupling
$g_s$, $g_0^2 = 8\pi g_s e^\Phi$, because
\bea
\frac{8\pi^2}{g_1^2} + \frac{8\pi^2}{g_2^2} =\frac{2\pi}{g_s e^\Phi}
~.
\eea
The duality cascade happens when one of coupling, for example $g_1$,
becomes large. We denote such a coupling to be $g_{SD}$, which we expect to be
of order one. The corresponding scale is
\bea
\mu_{SD} = \Lambda e^{\frac{8\pi^2}{3M} ( \frac{1}{g_0^2} -
\frac{1}{g_{SD}^2})} ~.
\eea
The scale where one of the coupling hits infinity is
\bea
\mu_{1\infty} = \Lambda e^{\frac{8\pi^2}{3M g_0^2}} ~, \\
\mu_{2\infty} = \Lambda e^{-\frac{8\pi^2}{3M g_0^2}} ~.
\eea
So $\mu_\infty$ is the location of the transition wall, and
$\mu_{SD}$ characterize its thickness.
Therefore the width of the step is
\bea
\Delta \mu_{SD} = 2 |\mu_{1\infty} - \mu_{SD}| ~,
\eea
and the separation between the two steps is
\bea
|\mu_{1\infty} - \mu_{2\infty}| ~.
\eea
Their ratio is
\bea
\frac{\Delta \mu_{SD}}{|\mu_{1\infty} - \mu_{2\infty}|}
\approx \frac{8\pi^2}{g_{SD}^2} g_s e^\Phi ~.
\eea
In terms of the distance $r$, the separation between the step $l$ and
$l+1$ is $2 \pi r_l /3g_s M$. So the width of the step is
\bea
\Delta r_{SD} \approx \frac{16\pi^2}{3g_{SD}^2} \frac{r_l}{M} ~.
\eea

 In IR DBI inflation it is also convenient to write such a width in
terms of the e-folds that brane spends crossing the width of the step.
From the attractor solution $r\approx R^2/|t|$ and the fact that the
speed-limit does not change much across the step, we get
\bea
\Delta N_e^{SD} \approx \frac{16\pi^2}{3g_{SD}^2 }
\frac{N_e}{M} ~,
\label{NeSD}
\eea
where $N_e$ is the e-fold to the end of DBI inflation at the location of
the step.

\subsection{The width from multiple brane spreading}
\label{Sec:Spread}
In the multi-throat brane inflation scenario \cite{Chen:2004gc}, 
inflaton branes are
almost always generated in a large number.
A typically generation mechanism can be the flux-antibrane
annihilation.
Once they are
created at the end of the annihilation process, they can have
different velocities. This introduces a spread in the mobile brane
position. In this subsection, we estimate such a width. We express it
in terms of e-folds $\Delta N_e^{\rm spread}$.

The time-delay between the fastest brane and the static brane is the
time period that the potential accelerates the brane from $\dot
\phi=0$ to the speed-limit $\phi^2/\sqrt{N}$.
Such an acceleration can be approximately described by
\bea
\ddot \phi = V'(\phi) = m^2 \phi ~.
\eea
Consider initially, at $t=0$, $\dot \phi_{min}=0$ and $\phi_{min} =
\sqrt{T_3} r_{min} = \sqrt{N} h_{min} / R$.
At $t=\Delta t$, $\dot \phi = \phi^2/\sqrt{N}$. Here $N$ and $R$ are
the charge and scale of the throat, respectively.
We can therefore solve for $\Delta t$ assuming $m\Delta t \ll 1$.
Further, using $m\sim H$,
we get
\bea
\Delta N_e^{\rm spread} =H\Delta t
\approx \frac{h_{min}}{H R} ~.
\eea
For a long throat, the minimum warp factor is typically determined by
the Hubble deformation as a result of the dS space back-reaction. In
Ref.~\cite{Chen:2005ad,Chen:2006ni},
it is estimated to be $h_{min} \sim HR/\sqrt{N}$. So
\bea
\Delta N_e^{\rm spread} \sim \frac{1}{\sqrt{N}} ~.
\label{Nespread}
\eea
Comparing (\ref{NeSD}) and (\ref{Nespread}), we see that generically
$\Delta N_e^{SD} > \Delta N_e^{\rm spread}$,
although the relation can change for special parameters.
The common feature is that both of them are almost invariant
for different steps.

\end{document}